\documentclass[sigconf]{acmart} 

\usepackage{multirow}
\newcommand{\commentout}[1]{}
\newcommand{\mmm}[1]{\begin{tabular}{c}#1\end{tabular}}
\usepackage{enumitem}
\setlist[description]{leftmargin=\parindent,labelindent=0pt}
\setlist[itemize]{leftmargin=\parindent,labelindent=0pt}
\setlist[enumerate]{leftmargin=\parindent,labelindent=0pt}

\newcommand{\add}[1]{#1}

\AtBeginDocument{%
  }

\setcopyright{acmcopyright}
\copyrightyear{2023}
\acmYear{2023}
\setcopyright{acmlicensed}\acmConference[CHI '23]{Proceedings of the 2023 CHI Conference on Human Factors in Computing Systems}{April 23--28, 2023}{Hamburg, Germany}
\acmBooktitle{Proceedings of the 2023 CHI Conference on Human Factors in Computing Systems (CHI '23), April 23--28, 2023, Hamburg, Germany}
\acmPrice{15.00}
\acmDOI{10.1145/3544548.3580706}
\acmISBN{978-1-4503-9421-5/23/04}

\begin{document}

\title{WESPER: Zero-shot and Realtime Whisper to Normal Voice Conversion for Whisper-based Speech Interactions}

\author{Jun Rekimoto}
\orcid{0000-0002-3629-2514}
\affiliation{%
 \institution{The University of Tokyo}
\city{7-3-1, Hongo, Bunkyo-ku}
\state{Tokyo}
\country{Japan}
\postcode{113-0033}
}
\affiliation{%
\institution{Sony Computer Science Laboratories, Kyoto}
\city{13-1 Hontoro-cho, Shimogyo-ku, Kyoto-shi}
\state{Kyoto}
\country{Japan}
\postcode{600-8086}
}
\email{rekimoto@acm.org}

\renewcommand{\shortauthors}{J.Rekimoto}

\begin{abstract}
Recognizing whispered speech and converting it to normal speech creates many possibilities for speech interaction. Because the sound pressure of whispered speech is significantly lower than that of normal speech, it can be used as a semi-silent speech interaction in public places without being audible to others. Converting whispers to normal speech also improves the speech quality for people with speech or hearing impairments. However, conventional speech conversion techniques do not provide sufficient conversion quality or require speaker-dependent datasets consisting of pairs of whispered and normal speech utterances. To address these problems, we propose WESPER, a zero-shot, real-time whisper-to-normal speech conversion mechanism based on self-supervised learning. 
WESPER consists of a speech-to-unit (STU) encoder, which generates hidden speech units common to both whispered and normal speech, and a unit-to-speech (UTS) decoder, which reconstructs speech from the encoded speech units. Unlike the existing methods, this conversion is user-independent and does not require a paired dataset for whispered and normal speech. The UTS decoder can reconstruct speech in any target speaker's voice from speech units, and it requires only an unlabeled target speaker's speech data. We confirmed that the quality of the speech converted from a whisper was improved while preserving its natural prosody. Additionally, we confirmed the effectiveness of the proposed approach to perform speech reconstruction for people with speech or hearing disabilities.
\end{abstract}

\ccsdesc[100]{Human-centered computing~Sound-based input / output}
\ccsdesc[500]{Computing methodologies~Neural networks}
\ccsdesc[300]{Human-centered computing~Interface design prototyping}
\ccsdesc[300]{Human-centered computing~Mobile devices}


\keywords{speech interaction, whispered voice, whispered voice conversion, silent speech, artificial intelligence, neural networks, self-supervised learning}

\begin{teaserfigure}
  \centering
  \includegraphics[width=0.83\textwidth]{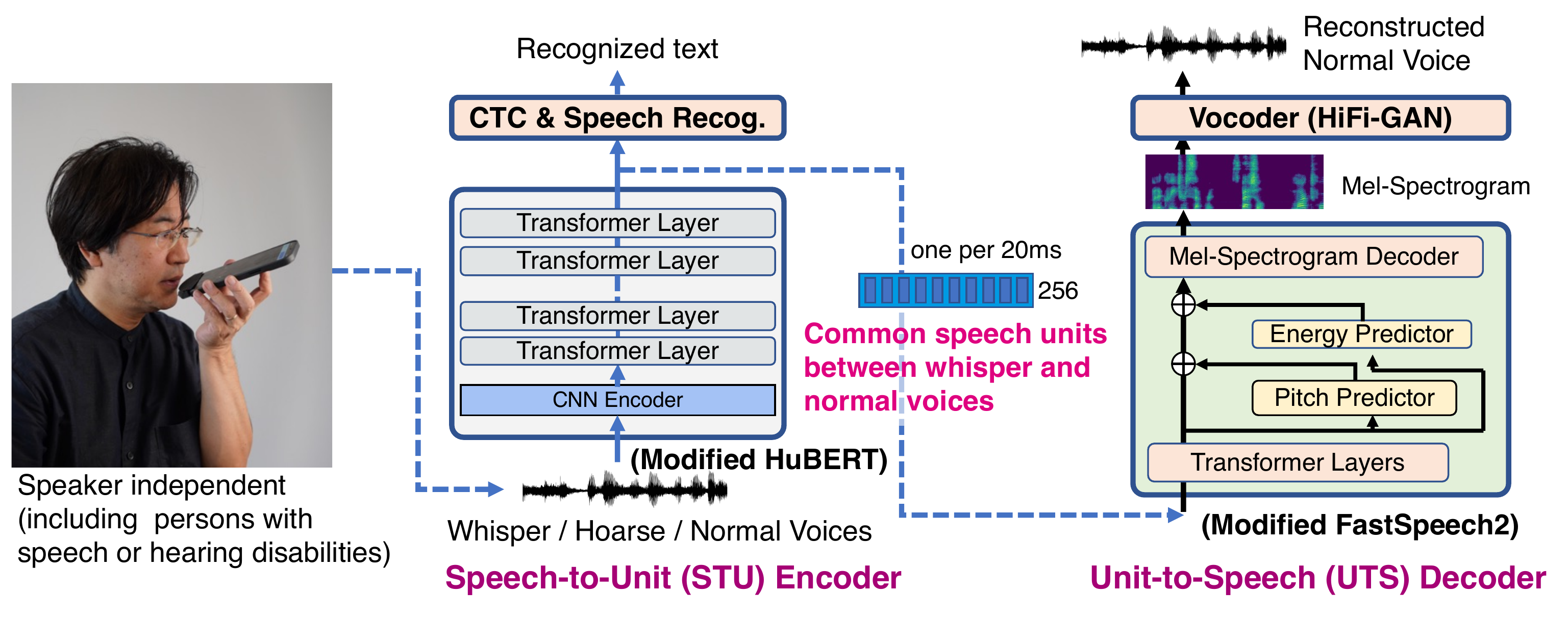}
  \caption{WESPER is a real-time whisper-to-normal speech conversion mechanism consisting of a speech-to-unit (STU) encoder that generates common speech units for whispered and normal utterances using self-supervised pre-training, and a unit-to-speech (UTS) decoder that recovers speech from the speech units. It achieves user-independent voice conversion in real time.}
    \Description{WESPER is a real-time whisper-to-normal speech conversion mechanism consisting of a speech-to-unit (STU) encoder that generates common speech units for whispered and normal utterances using self-supervised pre-training, and a unit-to-speech (UTS) decoder that recovers speech from the speech units. It achieves user-independent voice conversion in real time.}
  \label{fig:teaser}
\end{teaserfigure}

\maketitle

\section{Introduction}
Although voice interaction systems have been widely deployed, they are typically not easy to use in the presence of other people. Using voice commands in public places may be socially unacceptable, and there is a risk of confidential information being leaked. In addition, speaking during a conference call can be uncomfortable for people in the vicinity and can compromise the confidentiality of a conversation. 

To overcome these problems, various silent speech input techniques have been developed~\cite{10.1145/3172944.3172977,10.1145/3290605.3300376,10.1145/3242587.3242599,7310970,10.1145/2971763.2971765,10.1145/2634317.2634322}; however, these methods require special sensors and have not achieved high accuracy in speech recognition, remaining instead at the level of recognizing predefined commands. Conversion of unconditioned silent speech into normal vocal utterances has also not been achieved.

Additionally, silent speech could be employed to capture utterances from people with speech or hearing impairments. However, owing to the above-mentioned limitations, existing methods cannot meet the requirements of practical accessibility aids. 

In contrast to silent speech, we focus on {\it whispered} speech. Whispers are sufficiently low in sound pressure to ensure confidentiality, and whispering is almost equivalent to silent speech. Whispering can be captured with an ordinary microphone and does not require any special sensor configuration. People with speech disorders can still speak in a whisper or with a hoarse voice, although their vocal organs may have been injured or extracted; thus, their voices might be recognized.

To address these issues, we propose WESPER, a real-time and zero-shot whisper-to-normal voice conversion method based on self-supervised learning. It is designed to perform speaker independent conversion of whispered speech to normal speech. 
There is no need for per-user training, and paired datasets of whispered and normal utterances are not required. 
The proposed architecture consists of a speech-to-unit (STU) encoder that is pre-trained with normal and whispered speeches and a unit-to-speech (UTS) decoder that generates a target voice from speech units (Figure~\ref{fig:teaser}). By pre-training with (unpaired) whisper and normal voices, the STU can be trained to reduce the difference between normal and whispered utterances and output a common speech unit. 

The UTS decoder can be trained from the speech data of a specific speaker (without accompanying text labels). If the person's voice is used, a conversion can be performed to restore the whispered voice to the person's normal voice, or even to another person's voice.

Because the encoder and decoder operate in a non-autoregressive manner, the entire system operates in real-time. Therefore, when applied to teleconferencing, for example, a conference participant can speak in a whispered voice, which is then converted in real time, and others can listen to the speech played back in a normal voice.

\begin{figure}
  \centering
  \includegraphics[width=\linewidth]{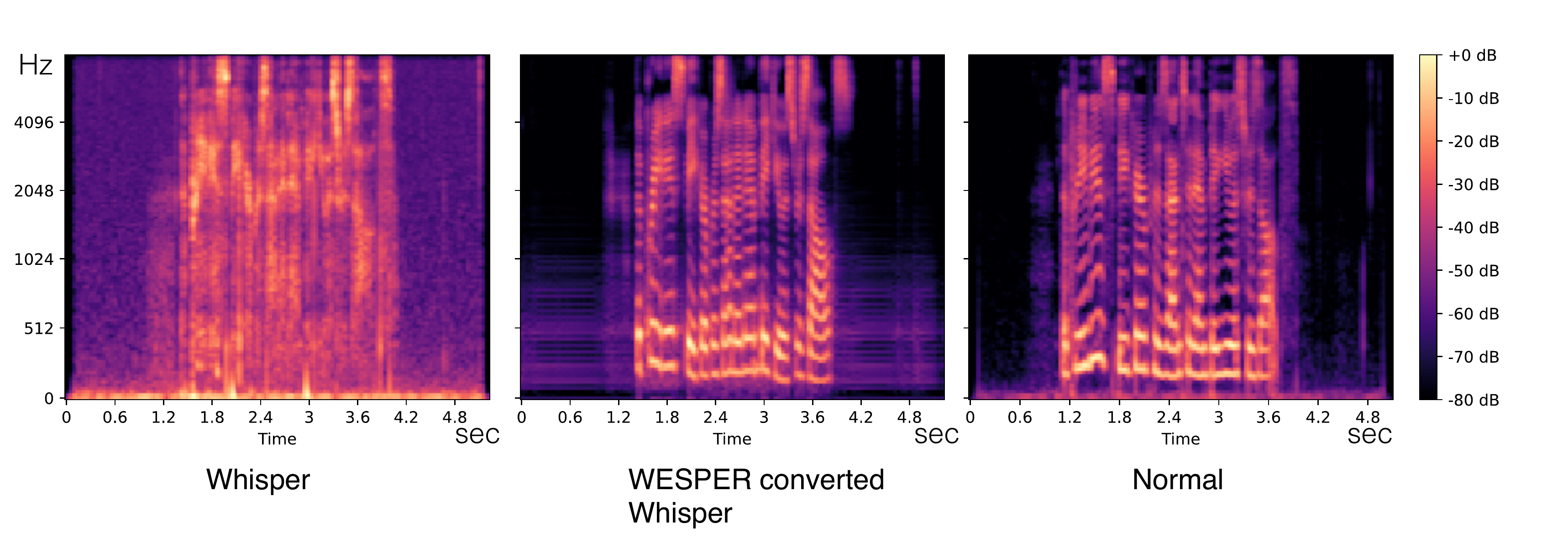}
    \caption{WESPER conversion result: Left: whispered speech; Middle: whispered speech converted by WESPER; Right: Normal speech by the same speaker, with the same transcription}
     \Description{WESPER conversion result: Left: whispered speech; Middle: whispered speech converted by WESPER; Right: Normal speech by the same speaker, with the same transcription}
  \label{fig:conversion}
\end{figure}

Figure~\ref{fig:conversion} shows mel-spectrogram examples of whisper-to-normal voice conversion using the proposed method. More conversion results are demonstrated in the accompanying video.

The contributions of this study can be summarized as follows.

\begin{itemize}
\item We propose a real-time, speaker-independent, vocabulary-free whisper-to-normal speech conversion method that can be trained only on unpaired whispers and normal speech.
\item The target voice can be learned by voice samples of a specific speaker without text transcription.
\item We experimentally confirmed an improvement in performance for normal speakers and for dysarthric or hearing-impaired speakers.
\end{itemize}

\begin{table*}
    \centering
    \begin{tabular}{c|cccc}
    \toprule
    model & training data & per-user dataset & \mmm{conversion to \\normal voice} & \mmm{(whisper)\\ speech recognition} \\
    \midrule
    \midrule
    \mmm{silent speech\\
    ex.\cite{10.1145/3172944.3172977,10.1145/3290605.3300376,10.1145/3242587.3242599,7310970,10.1145/2971763.2971765,10.1145/2634317.2634322}} & \mmm{special paired-dataset for\\dedicated sensors} & required &  both & \mmm{YES\\(commands, or characters)} \\ \hline
    Parotoron~\cite{parrotron} & paired whisper-normal voice & required & YES & \mmm{YES \\(through voice conversion)}\\ \hline
    DualVoice~\cite{10.1145/3526113.3545685} & \mmm{labeled, unpaired \\whisper and normal voice} & required & NO  & YES \\ \hline
    SilentVoice~\cite{silentvoice} & \mmm{labeled\\silentvoice (custom)} & required & NO & YES \\ \hline

    CycleGAN-VC~\cite{https://doi.org/10.48550/arxiv.1711.11293} & \mmm{unpaired, unlabeled\\whisper and normal voice} & not required & YES & NO \\ \hline
    
    MSpeC-Net~\cite{9052966} & \mmm{paired \\whisper-normal voice} & required & YES & NO \\ \hline
    
    AGAN-W2SC~\cite{https://doi.org/10.48550/arxiv.2111.01342} & \mmm{paired \\whisper-normal voice} & required & YES & NO \\ \hline
    
    {\bf WESPER} (ours) & \mmm{unpaired and unlabeled\\ whisper and normal voice} & not required & YES & \mmm{YES \\(through voice conversion)} \\
    \bottomrule
    \end{tabular}
    \caption{Research on silent and whispered speech}
    \Description{Research on silent and whispered speech}
    \label{tab:comparison}
\end{table*}

\section{Related Work}

\subsection{Research on Silent and Whispered Speech}

Various studies on silent speech have developed methods to recognize a user's silent utterances or silent commands using various sensor configurations, including lip reading, EMG (Electromyography), and ultrasound~\cite{10.1145/3172944.3172977,10.1145/3290605.3300376,10.1145/3242587.3242599,7310970,10.1145/2971763.2971765,10.1145/2634317.2634322}.
However, these systems require a special dataset to perform training, which increases the need for special sensors and prevents these technologies from being widely used. Hence, these systems offer limited vocabulary (less than 100 commands are available, typically around 30). Although silent speech recognition has sometimes been identified as a possible approach to help people with dysphonia, these methods cannot interpret unrestricted free speech owing to the vocabulary limitation.

Whispered speech has some similar characteristics to silent speech, such as preserving social acceptability in public spaces. However, it can be more widely adopted, given that ordinary microphones can be used.

SilentVoice~\cite{silentvoice} is a speech interaction technique that uses {\it ingressive speech}, an utterance made while inhaling. The amplitude of ingressive speech is low, and it can be considered a variation of silent speech. It requires a microphone-like device placed very close to the mouth, and training is required for users to speak correctly with ingressive speech. A custom corpus of ingressive speech with text transcriptions is also required.

Research on whisper voice recognition~\cite{Denby:2010:SSI:1746726.1746804,Freitas:2016:ISS:3001610,10.1109/TASLP.2017.2738559,Chang2020-ks} has previously been conducted.
The Alexa smart speaker supports a {\it whisper mode}~\cite{ai-alexa}. In this mode,
when a user whispers to Alexa, Alexa responds in a whisper~\cite{Cotescu2019-fr}. 

DualVoice~\cite{10.1145/3526113.3545685} has also been proposed as a method to perform end-to-end recognition of whispered voices based on a self-supervised speech recognition system based on wav2vec2.0~\cite{wav2vec2} and HuBERT~\cite{10.1109/TASLP.2021.3122291}. They also proposed an interaction technique to distinguish between whispered and normal utterances. WESPER can be combined with DualVoice to selectively convert a user's whispered voice.

\subsection{Speech Conversion}

Normal speech conversion technologies have been developed to convert one voiced utterance to another~\cite{pmlr-v97-qian19c,hayashi_na_vc,StarGANVC,SEGAN,https://doi.org/10.48550/arxiv.1711.11293}; however, the conversion quality of these methods remains unsatisfactory when applied to whisper-to-normal voice conversion. 

Recently, several machine learning-based whisper-to-normal voice conversion techniques have been investigated~\cite{MLwhispsurvey}. To compare these techniques, it is important to consider the quality of conversion and the required characteristics of the dataset used for training. 
A {\it paired whisper-normal voice} dataset (i.e. a dataset containing both whispered and normal versions of the same utterance) or a {\it labeled whispered voice} dataset (i.e. a dataset containing whispered utterances accompanied by text labels) will require a significant amount of effort to prepare. Conversely, if an {\it un-paired and un-labeled whisper voice} dataset is used (i.e. a dataset containing only whisper utterances without corresponding normal versions or labels), the effort required to prepare the dataset is low.

Attention-guided generative adversarial network for whisper to normal speech conversion (AGAN-W2SC) is a GAN-based whisper to normal speech conversion~\cite{https://doi.org/10.48550/arxiv.2111.01342}.  
It converts a whispered voice represented as mel-spectrogram into the corresponding normal speech's mel-spectrogram.
It is based on GAN, and the attention-map is used for the conversion.  
It requires a paired whisper-normal speech dataset.
MSpeC-Net~\cite{9052966} is an autoencoder-based multi-domain voice conversion and supports whisper-to-normal conversion. It also requires a paired whisper-normal voice dataset.

Moreover, whispered speech can be converted to text using speech recognition and generate normal speech using text-to-speech methods~\cite{s2s_challenge}. 
However, this approach requires a labeled dataset for whispered speech recognition, and prosodic information contained in whispered utterances is lost because the intermediate text representation does not convey such information.

Parrotron~\cite{parrotron} is a speech conversion system designed to improve the speech of speakers with dysplasia. It is based on an encoder-decoder model that conforms to the text-to-speech system, Tacotron~\cite{DBLP:journals/corr/WangSSWWJYXCBLA17}; however, it differs from Tacotron in that both the input and output data are formatted as mel-spectrograms.
It also requires paired source and target speeches, making the construction of the required dataset relatively difficult.

Phonetic posteriorgrams (PPGs)~\cite{7552917} are intermediate information obtained from automatic speech recognition. PPGs can be used for many-to-one speaker conversion because they represent the articulation of spoken content speaker independently.
To our knowledge, the effects of PPGs on whispered voice have not been investigated. Our method also uses intermediated speech units, but it does not require text-based corpus as in the case of automatic speech recognition (ASR) and PPGs.

In contrast, our proposed system requires only samples of unpaired target and source speech, and it does not require accompanying text transcriptions, thus making datasets easy to prepare and independent of the target language.

The characteristics of related technologies are summarized in Table~\ref{tab:comparison}.
We also demonstrate whisper to normal conversion examples including NMSE-DiscoGAN~\cite{discoGAN}, MSpeC-Net~\cite{9052966}, CycleGAN-VC~\cite{https://doi.org/10.48550/arxiv.1711.11293}, AGAN-W2SC~\cite{https://doi.org/10.48550/arxiv.2111.01342}, and WESPER at \url{https://wesperproj.github.io/}.

\subsection{Self Supervised Representation Learning for Speech}

Recently, combining pre-training with self-supervised representation learning on unlabeled speech data and fine-tuning on labeled speech data has attracted attention. These systems are primarily intended for speech recognition applications, but they have also been applied to perform speaker, language, and emotion recognition~\cite{Pepino2021-hv,Yi2020-re}.

In particular, the pre-training method of hidden-unit BERT (huBERT)~\cite{10.1109/TASLP.2021.3122291} is similar to that of the masked language model used in bidirectional encoder representations from transformers (BERT)~\cite{BERT} in natural language processing. It is designed to mask part of the input and estimate the corresponding expression features from the rest of the input. With this pre-training, the model can learn the acoustic properties of the input data and the characteristics of the speech.

For use as ASR, after pre-training, fine-tuning is performed on only a small amount of the audio data with text transcriptions. A projection layer and a connectionist temporal classification (CTC) layer~\cite{ctc} has been added to generate text transcriptions from audio waveforms.

As reported in~\cite{wav2vec2,Yi2020-re}, self-supervised ASR achieved a speech recognition accuracy comparable to conventional state-of-the-art ASRs with fine-tuning only on a small amount of labeled speech data. Therefore, this architecture could be suitable for recognizing whispered voice recognition with limited whispered speech corpora.

WESPER is unique in that it uses pre-training to reduce the difference between the latent vectors encoding whispered and normal utterances.

\subsection{Textless-NLP}

Recently, text-free speech processing and speech conversion methods have been developed. Through self-supervised learning, these systems derive latent representations from speech data that is not accompanied by text transcriptions.
Textless-NLP\cite{gslm,textlessEmotion} and AudioLM~\cite{audioLM} do not use text transcriptions or phoneme symbols in speech processing systems; they use discrete units constructed by self-supervised learning. Soft discrete unit is another approach for textless speech processing~\cite{van_Niekerk_2022}. 

Our proposed method also uses non-discrete vectors as latent speech representations obtained from self-supervised learning and does not explicitly use text transcriptions or phoneme symbols. Our research is unique because we demonstrate that whispered and normal speech can be represented by similar speech units through self-supervised learning. In addition, our method is also designed to work seamlessly with a speech generation system in the later stages.

\section{The WESPER Voice Conversion Model}

WESPER consists of an STU encoder and a UTS decoder. The STU converts whispered or normal speech into common speech units. The UTS converts common speech units into mel-spectrograms that can be reconstructed as speech by a vocoder. WESPER is characterized by the fact that the common speech units of the same utterance can be similar, although the STU is only pretrained with (un-paired) whispered and normal speech, and it is not trained with a paired set of whispered and normal utterances. The details of each and the method used to train the model are described below.

\subsection{Speech-to-Unit (STU) Encoder}

\begin{figure}
  \centering
  \includegraphics[width=0.98\linewidth]{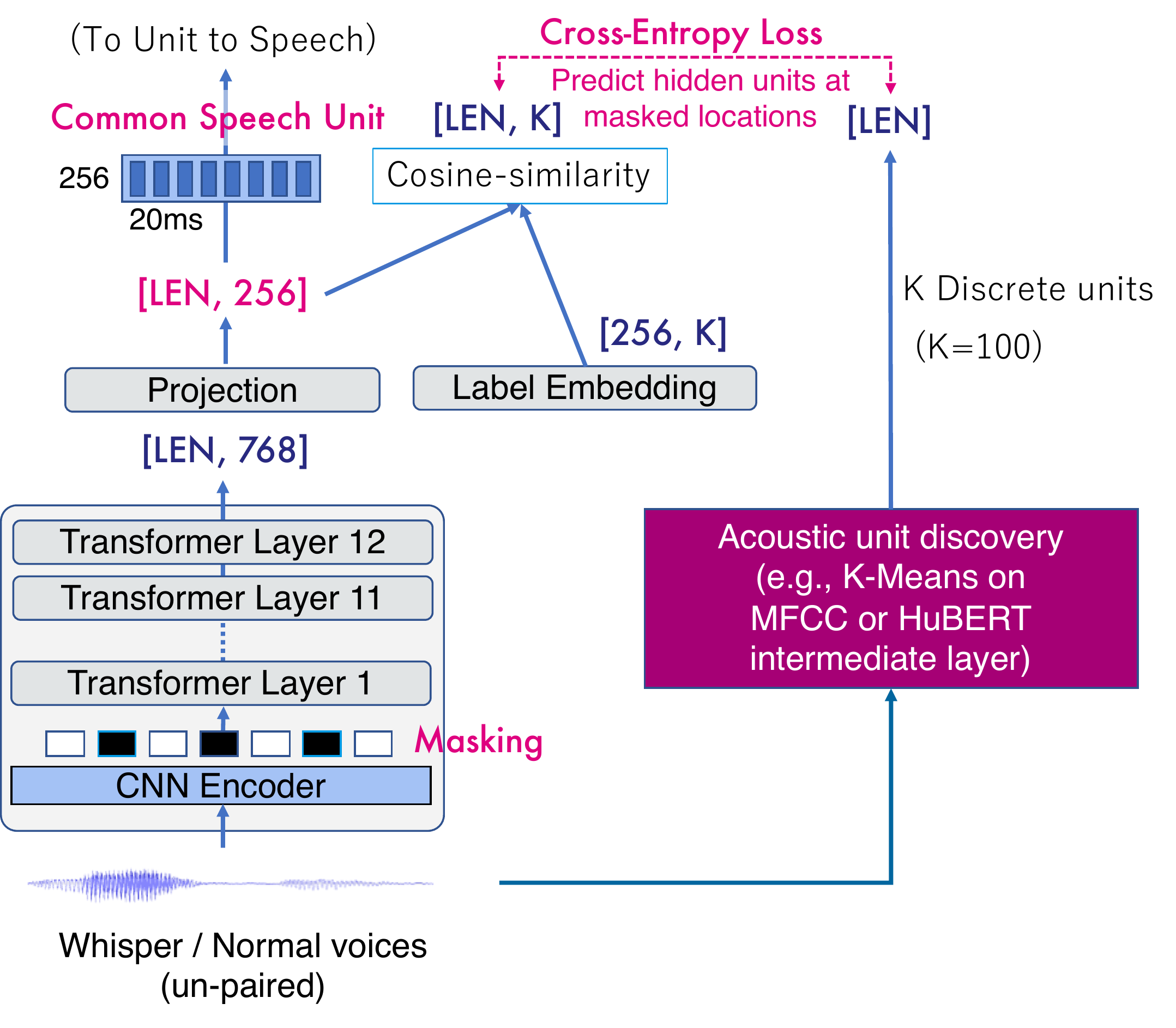}
    \caption{Overview of STU pre-training. Unpaired whispered or normal speech is used for pretraining. The transformer layer is trained by estimating discrete units from the masked input. The projection layer after the transformer layers generates 256-dimensional vectors (one every 20 ms), which are used as common speech units.}
   \Description{Overview of STU pre-training. Unpaired whispered or normal speech is used for pretraining. The transformer layer is trained by estimating discrete units from the masked input. The projection layer after the transformer layers generates 256-dimensional vectors (one every 20 ms), which are used as common speech units.}
  \label{fig:pretrain}
\end{figure}

\begin{figure}
  \centering
  \includegraphics[width=.95\linewidth]{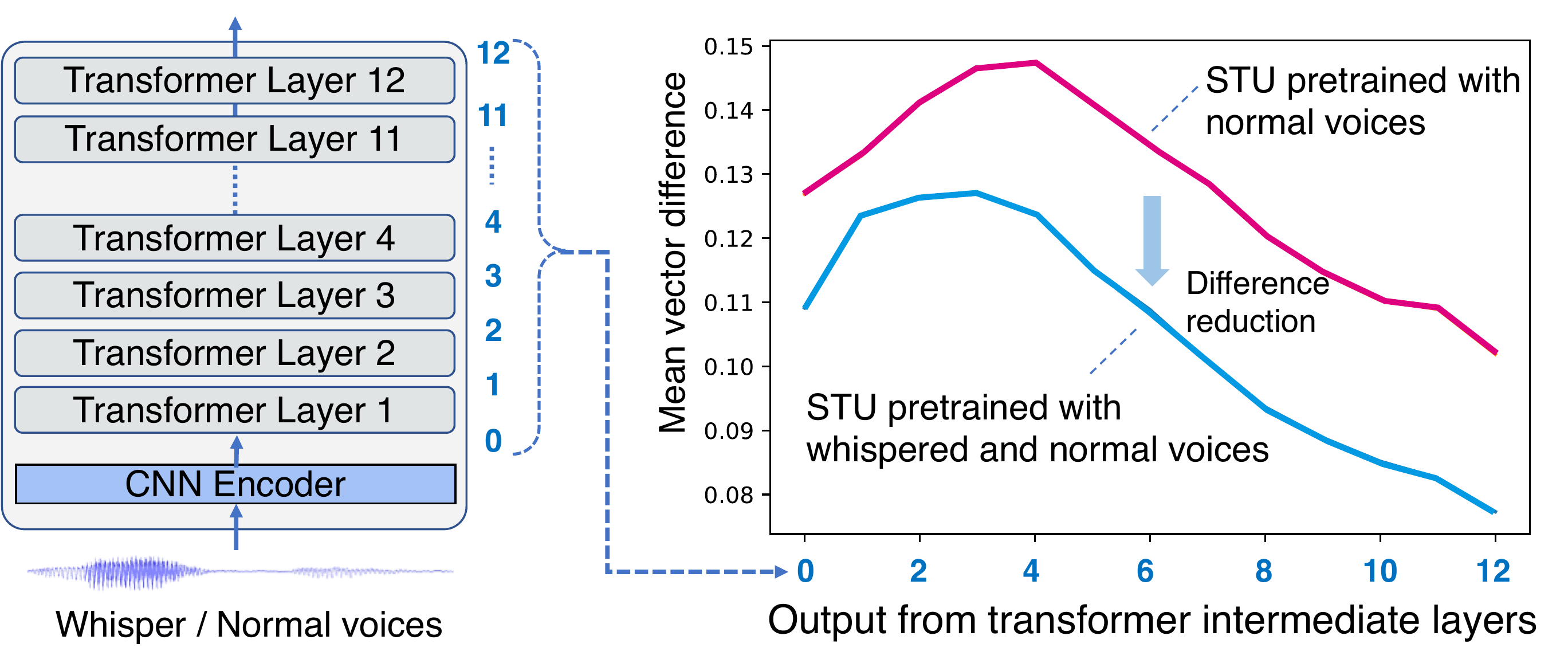}
  \includegraphics[width=.98\linewidth]{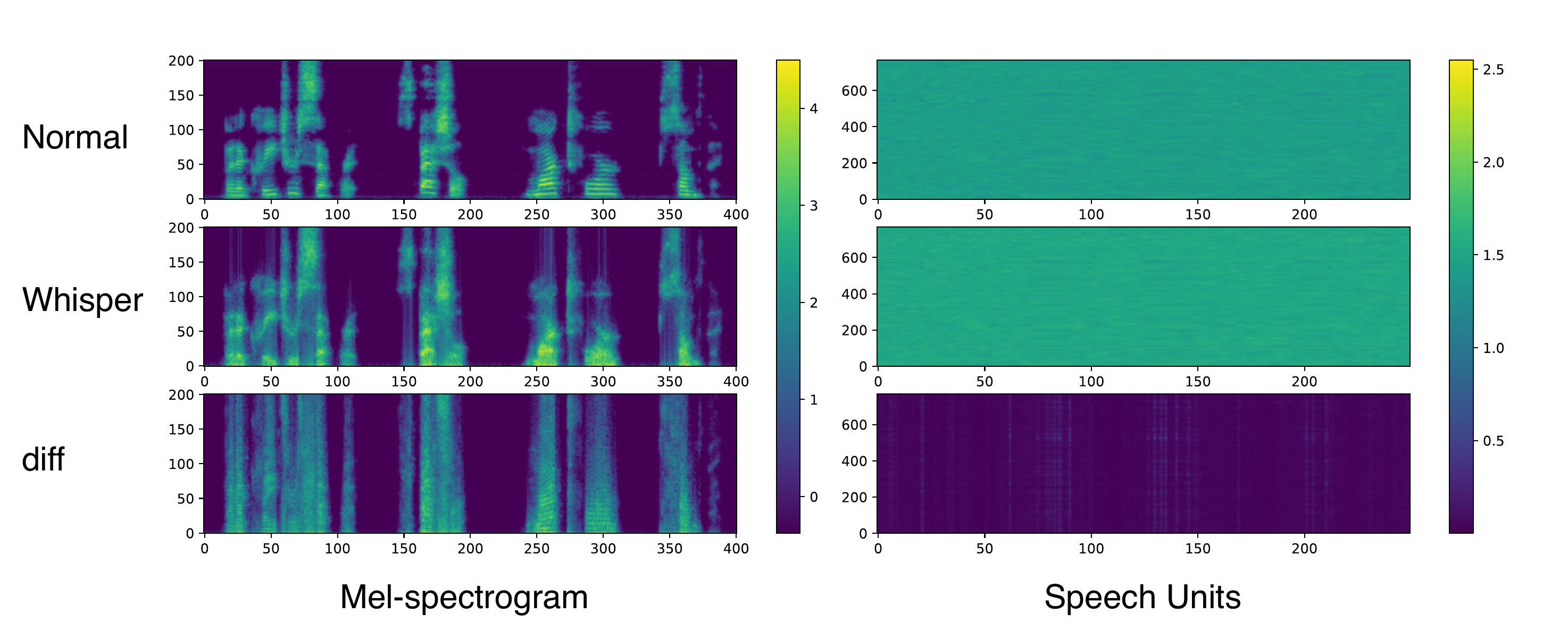}
  \includegraphics[width=.98\linewidth]{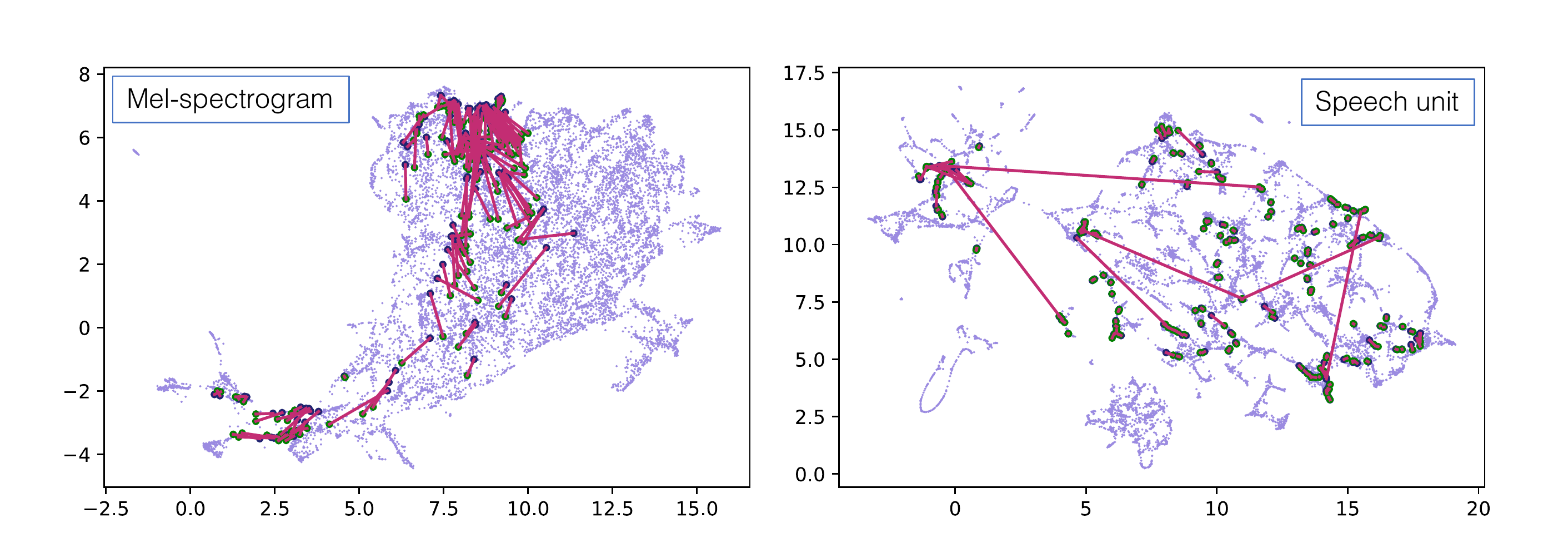}
  \caption{Comparison of whispered/normal voice differences: The comparison is made using normal speech and speech with whispering transformations. (Above) An STU pre-trained with normal and whispered speeches produced fewer differences than that pre-trained with normal speech only. As the transformer layer deepens, the difference between the two decreases. (Middle) There was a difference in the mel-spectrogram, which decreased with the speech units. (Bottom) UMAP~\cite{umap} visualizations of differences between whisper and normal voices in mel-spectrogram and speech-unit's space.}
    \Description{Comparison of whispered/normal voice differences: The comparison is made using normal speech and speech with whispering transformations. (Above) An STU pre-trained with normal and whispered speeches produced fewer differences than that pre-trained with normal speech only. As the transformer layer deepens, the difference between the two decreases. (Middle) There was a difference in the mel-spectrogram, which decreased with the speech units. (Bottom) UMAP~\cite{umap} visualizations of differences between whisper and normal voices in mel-spectrogram and speech-unit's space.}
  \label{fig:layer}
\end{figure}

The STU encoder takes audio waveform as the input and outputs units. It is based on HuBERT~\cite{10.1109/TASLP.2021.3122291}, a self-supervised neural network for speech, which pre-trains a large amount of unlabeled speech in a BERT~\cite{BERT}-like manner; it learns to recover the relevant parts from partially masked speech features and thereby acquires a speech-language model.

For our purpose, we want to generate speech units that are as identical as possible between whispered and normal speech.
To achieve this, we pre-trained an STU with a mixture of whispered and normal speech utterances; we used normal English speech with many speakers from the Librispeech 960h dataset~\cite{librispeech}. Librispeech speech data were mechanically converted to a whispered voice using an LPC-based audio conversion tool~\cite{towhisper}. Additionally, we used the wTIMIT speech dataset of normal and whispered speech~\cite{wTIMIT} with a speech length of 58 h.

Figure~\ref{fig:pretrain} shows the pre-training process of the STU in detail. Only the unpaired whispered or normal speech is used for pre-training. The transformer layer is trained by estimating discrete units from the masked input. Similar to HuBERT, the discrete target units are first generated by k-means clustering of the input speech data in the first stage, and then by k-means clustering of the outputs of the transformer intermediate layer.
In our experiment, we used 100 discrete units. The projection layer after the transformer layers generates a sequence of 256-dimensional vectors (one per 20 ms), which are used as common speech units.

In the STU, 12 transformer layers are placed in front of the CNN feature extractor (Figure~\ref{fig:teaser}). After pre-training, by comparing the output from each layer, we confirmed that (1) the difference in speech unit values between whispered and normal voices decreased with increasing layer depth; (2) the difference decreased when the STU was pre-trained with both whispered and normal speech utterances compared to pre-training with normal speech only (Figure~\ref{fig:layer}). Figure~\ref{fig:layer} (bottom) also shows the UMAP~\cite{umap} visualization of whispered and normal speech in mel-spectrogram space and common speech unit space. The corresponding periods of two speeches are connected by lines. As shown in the figure, the differences are reduced in the common speech unit space (except for a few long distances).

Although the speech feature values of whispered and normal voices are different, we assume this is addressed through the self-supervised pre-training so that utterances with similar linguistic standpoints are represented with similar units. 
We speculate that our learning method can extract common pronunciations from both normal and whispered utterances, similar to how self-supervised pre-training methods can extract common pronunciations from the utterances of different speakers in an ASR system.

In addition, we can fine-tune the STU with (transcript-attached) wTIMIT and Librispeech datasets by adding a CTC layer~\cite{ctc} as in~\cite{10.1145/3526113.3545685}; therefore, STU could be used as an ASR to recognize whispered and normal speeches.

\subsection{UTS Decoder}

The UTS decoder takes the speech units generated by the STU as the input and reconstructs the target (normal) speech. It is based on the non-autoregressive text-to-speech (TTS) system FastSpeech2~\cite{fastspeech2}. While the original FastSpeech2 includes an embedding layer that takes text or phoneme tokens and transforms them into a sequence of vector embeddings, we can eliminate this layer because UTS takes speech units as direct input. The original FastSpeech2 also includes a duration estimator that estimates the duration of each phoneme token and a length regulator that adjusts the number of internal vectors according to the estimated duration. These parts can also be eliminated for our purpose because the STU generates speech units at a constant rate.  

When learning the original FastSpeech2, it was necessary to provide the duration of each phoneme in the corpus as a ground truth by an external tool such as Montreal Forced Aligner~\cite{MFA}. Because of this limitation, FastSpeech2's learning is language-dependent. Conversely, UTS does not require duration estimation, and UTS can be language-independent.

The output of UTS is a mel-spectrogram, as in FastSpeech2. A vocoder (HiFi-GAN~\cite{hifigan}) converts this into an actual speech waveform.

\begin{figure}
  \centering
  \includegraphics[width=0.9\linewidth]{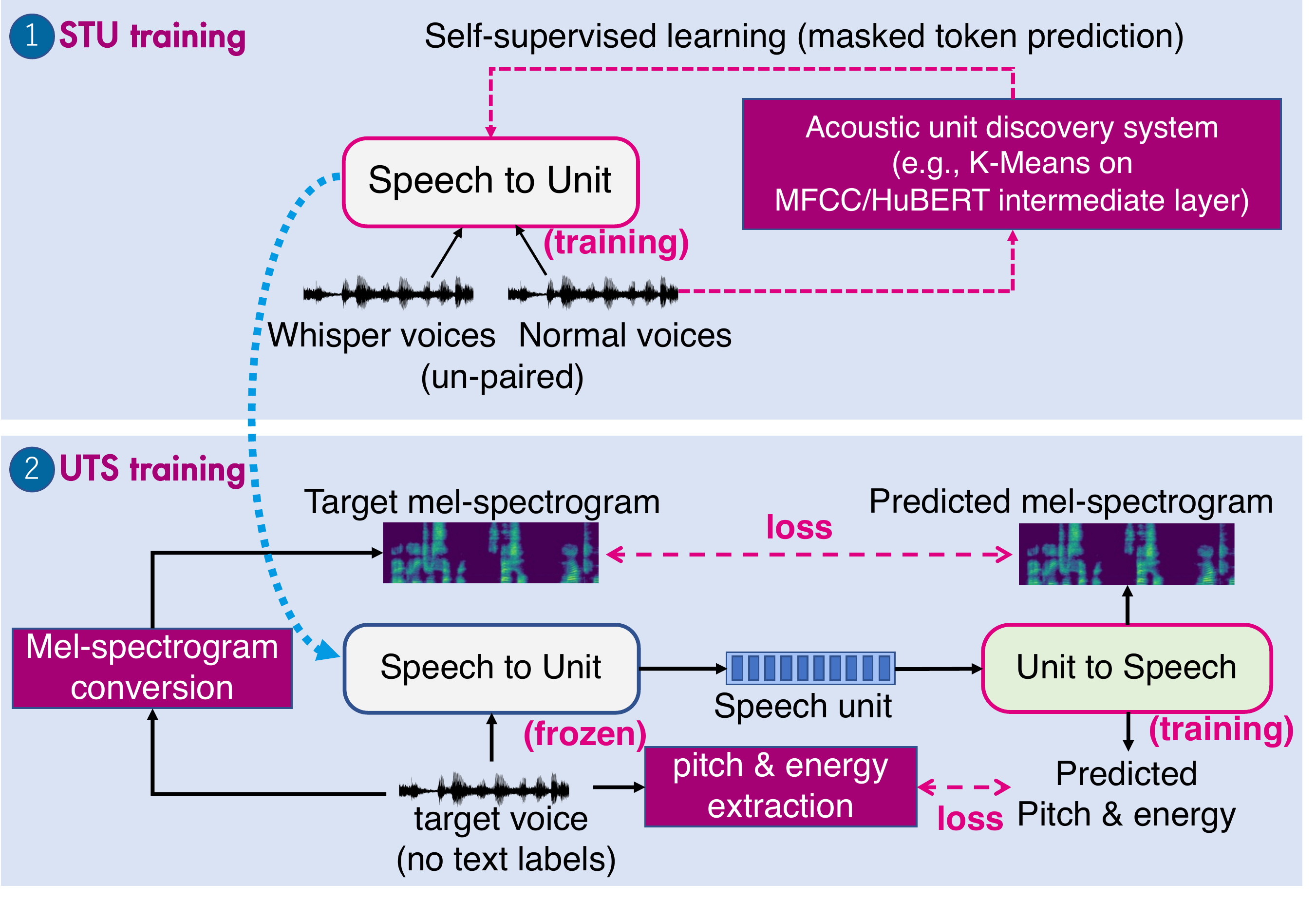}
  \caption{STU training (1) and UTS training (2): The UTS learns to decode speech units to the target speech using only wave data of the target voice with (frozen) STU. No text labeled dataset is required. Notably, WESPER was not trained on a labeled corpus, but only pre-trained on normal and whispered voices.}
  \Description{STU training (1) and UTS training (2): The UTS learns to decode speech units to the target speech using only wave data of the target voice with (frozen) STU. No text labeled dataset is required. Notably, WESPER was not trained on a labeled corpus, but only pre-trained on normal and whispered voices.}
  \label{fig:UTS}
\end{figure}

In a regular TTS system, the target speech and the corresponding text labels are needed for training. By contrast, the proposed UTS requires only the target speech and no text labels. The target speech is passed through the STU to obtain a sequence of speech units associated with the speech waveform and used to train the UTS (Figure~\ref{fig:UTS}). In our experiment, the UTS was trained using speech data from a single speaker of LJSpeech~\cite{ljspeech17}, and data from other speakers taken from narration data.

\begin{figure}
  \centering
  \includegraphics[width=0.95\linewidth]{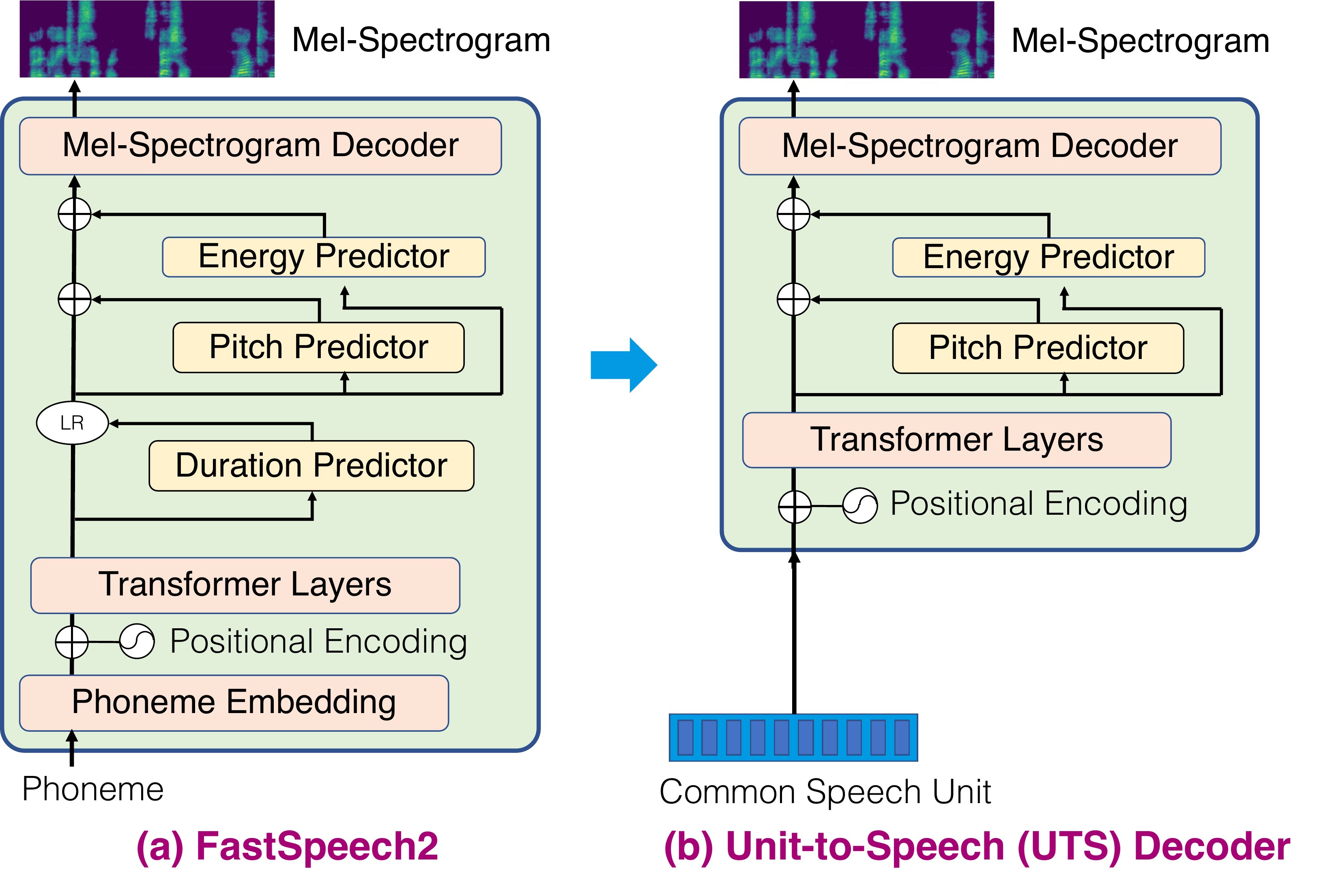}
    \caption{Comparison of FastSpeech2~\cite{fastspeech2} and UTS decoders: FastSpeech2 needs to predict the duration of each phoneme, whereas UTS accepts a common speech unit with the same duration. This eliminates the duration predictor and the length regulator (LR in the figure). Phoneme embedding can also be eliminated because the common speech unit does not consist of discrete tokens.}
    \Description{Comparison of FastSpeech2~\cite{fastspeech2} and UTS decoders: FastSpeech2 needs to predict the duration of each phoneme, whereas UTS accepts a common speech unit with the same duration. This eliminates the duration predictor and the length regulator (LR in the figure). Phoneme embedding can also be eliminated because the common speech unit does not consist of discrete tokens.}
  \label{fig:FS2}
\end{figure}

\section{System Configuration}

\begin{figure}
    \centering
    \includegraphics[width=0.9\linewidth]{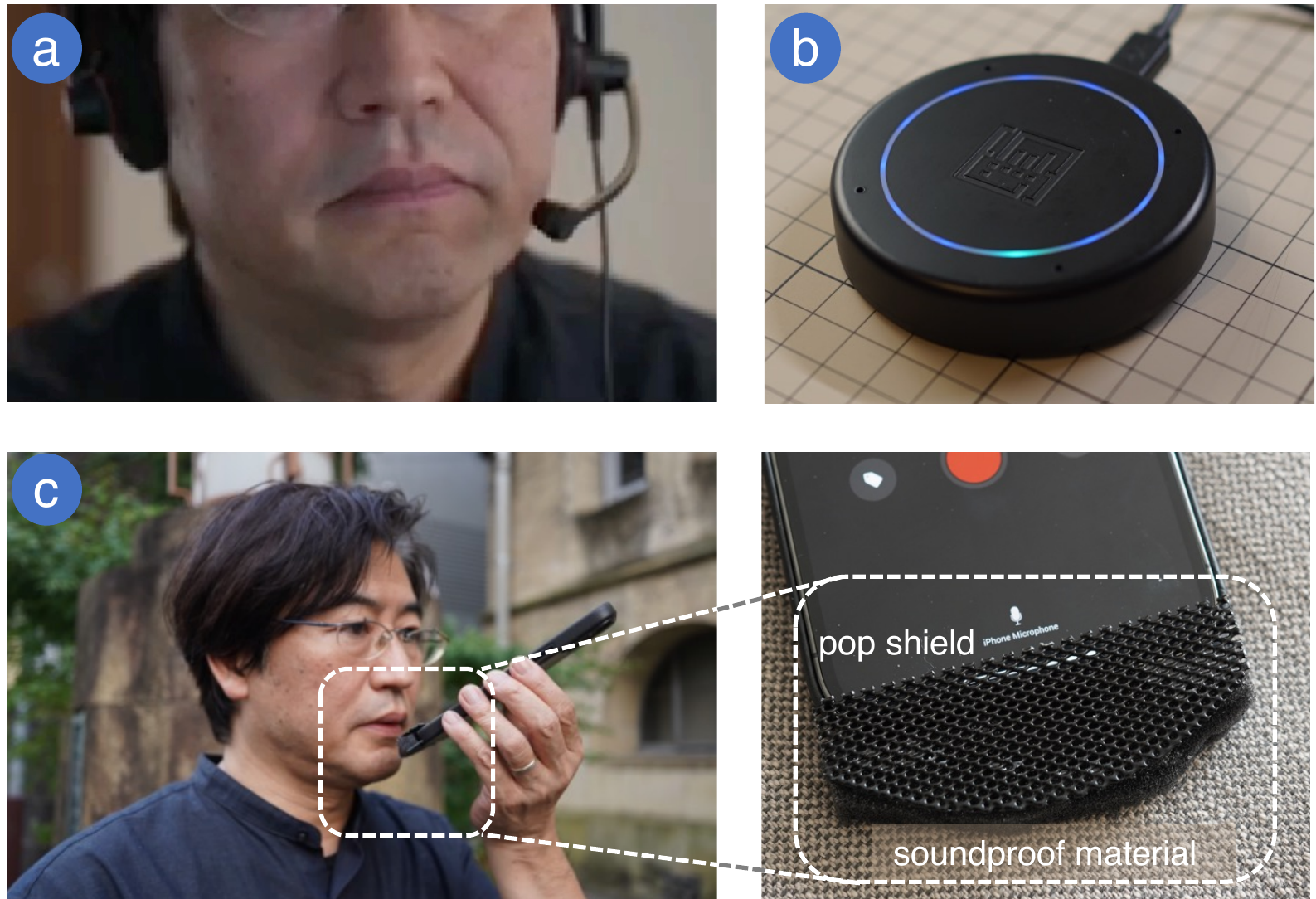}
    \caption{WESPER speech input device: (a) headset, (b) array (directional) microphone, and (c) cell phone microphone with pop-protection and soundproofing material. }
   \Description{WESPER speech input device: (a) headset, (b) array (directional) microphone, and (c) cell phone microphone with pop-protection and soundproofing material. }
    \label{fig:device}
\end{figure}

\begin{figure*}
    \centering
    \includegraphics[width=\textwidth]{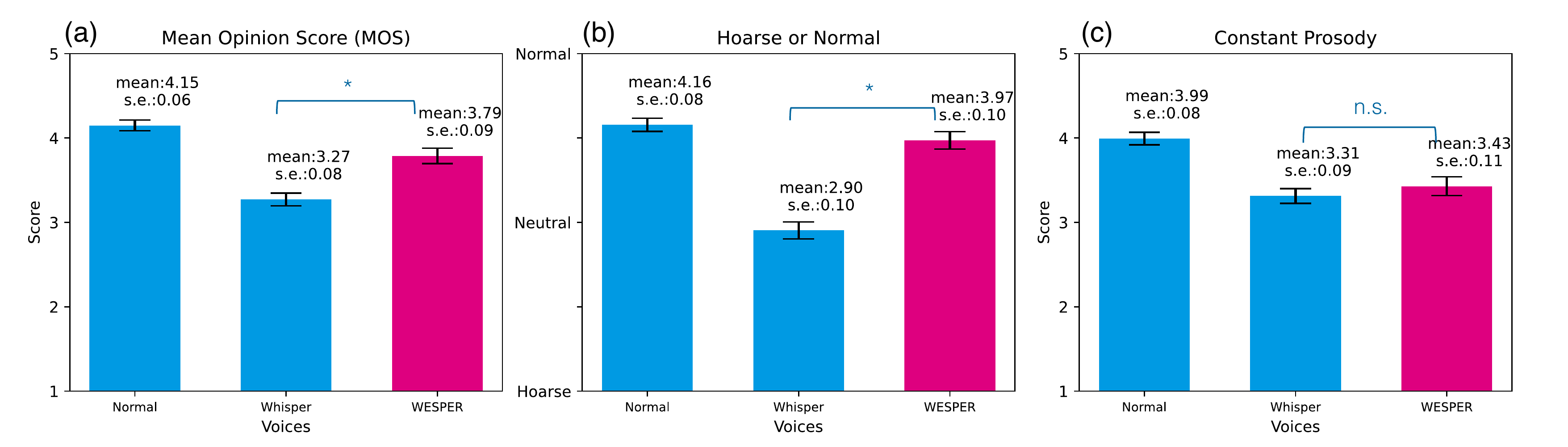}
    \caption{Quality of whisper-to-normal conversion: (a) Mean Opinion Scores (MOS) of normal, whispered, and WESPER-converted whispered voices. (b) Hoarse-normal voice rating, and (c) natural prosody rating ({\tt s.e.}: standard error, {\tt *}: $p < 0.01$ by t-test, {\tt n.s.}: not significant)}
   \Description{Quality of whisper-to-normal conversion: (a) Mean Opinion Scores (MOS) of normal, whispered, and WESPER-converted whispered voices. (b) Hoarse-normal voice rating, and (c) natural prosody rating ({\tt s.e.}: standard error, {\tt *}: $p < 0.01$ by t-test, {\tt n.s.}: not significant)}
    \label{fig:MOS}
\end{figure*}

We designed the WESPER model using the PyTorch framework. The STU is based on HuBERT, and the UTS is based on a modified implementation of FastSpeech2 PyTorch implementation by~\cite{chien2021investigating}. We used Librispeech and wTIMIT (both of which include normal and whispered speech) for pre-training. With a dual NVIDIA R6000, pre-training took 48 h. UTS training took 26 h for each target voice. 

The total processing time required to perform the conversion was approximately 1/20th of the actual speech duration on a single NVIDIA R6000 and 1/10th on an Apple M1 Max CPU. The actual quality of the conversion is demonstrated in the attached video.

We designed two types of interfaces. The first operated in a push-to-talk style, where the user first speaks in a whispered or normal voice while pressing a button. Thereafter, when the button is released, the speech waveform recorded during that time is sent to the speech-conversion neural networks, and the result is immediately played back. The other detected a non-audio period of input speech and automatically converted speech segments without additional user input.

Figure~\ref{fig:device} shows the current WESPER speech input configurations. We tested with (a) a headset, (b) directional microphone with four arrayed MEMS (Microelectromechanical systems)  microphones~\cite{respeaker}, and (c) mobile phone microphone with pop-guard to avoid whispering pop noise and soundproofing material to reduce ambient noise.

\section{Evaluation}

The WESPER mechanism allows whisper-to-normal conversion independent of the input speaker. Here, we evaluate the conversion quality in three aspects, considering whisper-to-normal conversion and voice reconstruction for people with speech and hearing disorders.

\subsection{Quality of Whisper-to-Normal Conversion}

\begin{table*} 
\commentout{
\begin{tabular}{l|rr}\toprule
 test & WER (\%) & CER (\%) \\\midrule
 wTIMIT(N) & 11.55 & 4.66 \\
                        wTIMIT(W) & 44.70 & 28.38 \\\midrule
 wTIMIT(N)+ WESPER & 13.44 & 5.53 \\
  wTIMIT(W) + WESPER & {\bf 29.37} & {\bf 15.01} \\
\bottomrule
\end{tabular}
\begin{tabular}{rl}
wTIMIT(N) : & wTIMIT normalvoice \\
wTIMIT(W) : & wTIMIT whispered voice \\
\end{tabular}
}
\vbox{
\begin{tabular}{c|ll|rrr}\toprule
model & train (finetune) &  test & WER (\%) & CER (\%) & BLEU \\\midrule
\multirow{3}{*}{Google} & \multirow{3}{*}{(trained by Google)} & wTIMIT(N) & 11.55 & 4.66 & 0.76 \\
            & &             wTIMIT(W) & 44.70 & 28.38 & 0.34 \\
& &  {\bf WESPER[wTIMIT(W)}] & {\bf 26.68} & {\bf 12.70} & {\bf 0.52}\\
\midrule
\multirow{3}{*}{\mmm{HuBERT\\base}} & Librispeech & wTIMIT(N) & 21.06 & 8.17 & 0.54 \\
& Librispeech & wTIMIT(W) & 33.06 & 15.45 & 0.38\\
& librispeech + wTIMIT(N,W) & wTIMIT(W) & {\bf 13.75} & {\bf 5.47} & {\bf 0.70}\\
\bottomrule
\end{tabular}

\begin{tabular}{rlcrl}
wTIMIT(N): & wTIMIT~\cite{wTIMIT} normal voice & & wTIMIT(W): & wTIMIT whisper voice \\
Google: & Google Cloud Speech-to-Text~\cite{googlespeech} & & WESPER[$\bullet$]: & WESPER converted $\bullet$ \\
HuBERT: & \multicolumn{4}{l}{HuBERT base model, pretrained with Librispeech~\cite{librispeech} + wTIMIT(N,W) } \\
\end{tabular}
}
\caption{Whispered voice recognition accuracy of Google Cloud Speech-to-Text~\cite{googlespeech} as a reference ASR: The recognition rate of whispered voice in normal ASR was not high; however, when the result of converting whispered voice to normal voice using WESPER was recognized, the recognition rate improved. Notably, WESPER was not trained on labeled data, but it was pre-trained on (un-labeled and unpaired) normal and whispered voices.}
\Description{Whispered voice recognition accuracy of Google Cloud Speech-to-Text~\cite{googlespeech} as a reference ASR: The recognition rate of whispered voice in normal ASR was not high; however, when the result of converting whispered voice to normal voice using WESPER was recognized, the recognition rate improved. Notably, WESPER was not trained on labeled data, but it was pre-trained on (un-labeled and unpaired) normal and whispered voices.}
\label{tab:wer}
\Description[Whispered voice recognition accuracy]{Whisper voice recognition accuracy}
\end{table*}

To evaluate the quality of the converted speech, we recruited 50 gender-balanced participants online using the Prolific crowdsourcing system~\cite{prolific}, all of whom were over 18 and fluent in English. Each participant listened to four sets of normal speech, whispered speech, and WESPER-converted whispered speech (12 voices in total) with a web-based user interface and rated the utterances on a 5-point Mean Opinion Score (MOS) and other questionnaires (examples of voices are included in an attached video). 

We used the LJSpeech-trained WESPER voice as the target voice and the same transcription sentence for all voices to avoid differences in impressions based on sentence content. 

The results are presented in Figure~\ref{fig:MOS}.
Figure~\ref{fig:MOS} (a) shows the results of the MOS evaluation. The WESPER-converted voice showed a score between that of normal and whispered voices. It was confirmed that WESPER conversion improved the MOS of the original whispered speech ($p < 0.01$ by pairwise t-test, Cohen's d effect size = 0.54).
Figure~\ref{fig:MOS} (b) shows the responses to the question ``Is this voice hoarse or normal?'' and there is a clear improvement in the voices converted with WESPER ($p < 0.01$).
Figure~\ref{fig:MOS} (c) shows the responses to the question ``Is the voice using consistent articulation, standard intonation, and prosody?'' Here, WESPER and whisper showed almost equal results. Notably, WESPER did not affect the naturalness of the prosody.
Considering the speech converted by WESPER is generated by the unit-to-speech module and the vocoder, this result indicates that WESPER can preserve the natural prosody of the original speech.

\subsubsection*{Multiple Stimuli with Hidden Reference and Anchor 
 (MUSHRA) evaluation}

\begin{figure}
    \centering
    \includegraphics[width=0.8\linewidth]{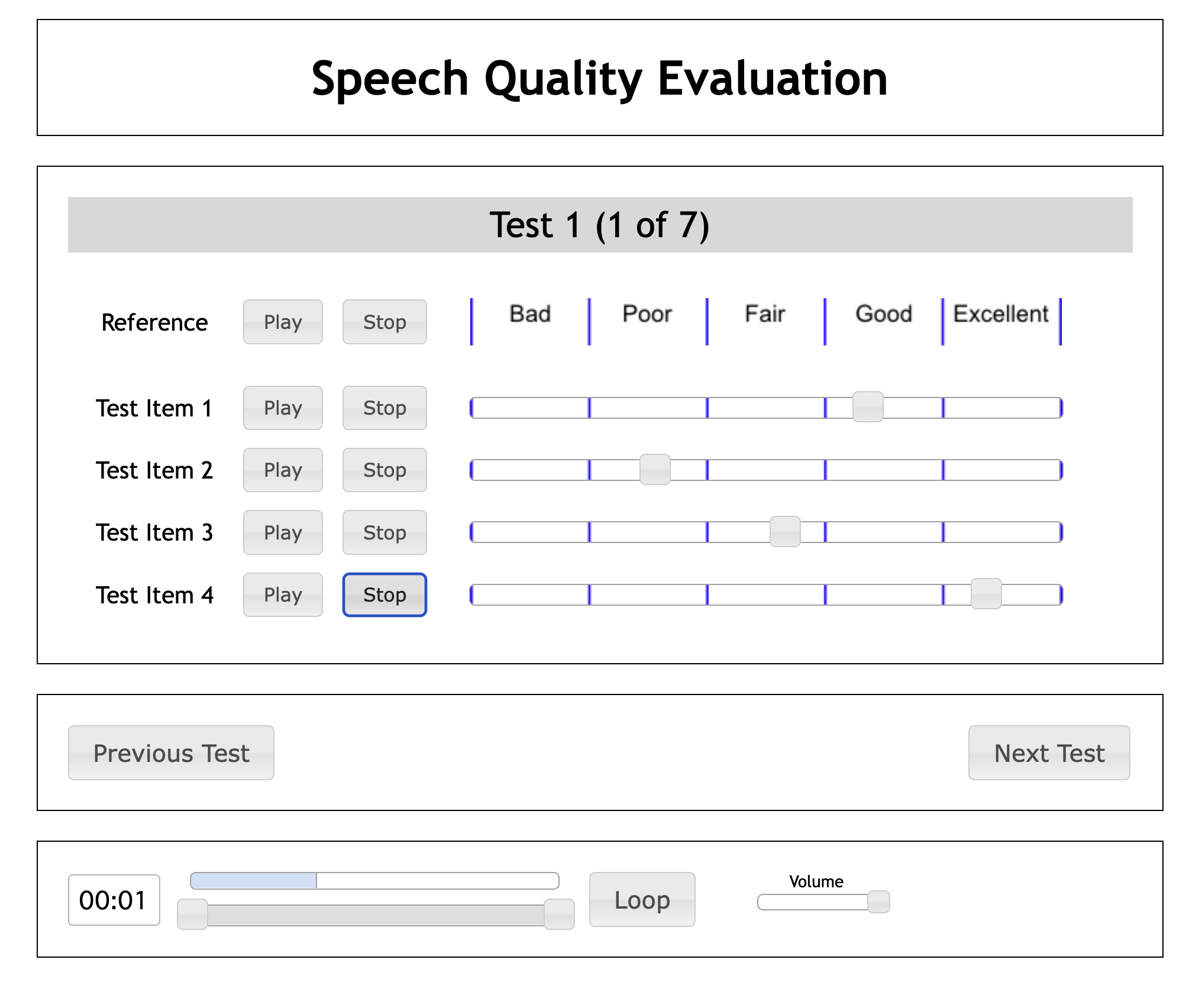}
    \caption{\add{An example of MUSHRA evaluation web interfaces.}}
    \Description{An example of MUSHRA evaluation web interfaces.}
    \label{fig:mushraweb}
\end{figure}

In addition to the MOS test, we evaluated speech quality using Multiple Stimuli with Hidden Reference and Anchor (MUSHRA). MUSHRA is a method for evaluating the perceived quality of audio as defined by ITU-R Recommendation BS.1534-3~\cite{mushra}.
MUSHRA uses anchor audio and other audios, and an evaluator is expected to assign ratings (from 0 to 100) to the audio by comparing it to the anchor audio as the reference. \add{For each rating, participants can play each voice as many times as they like. } Because of the presence of anchor and hidden reference audio, MUSHRA is considered more reliable than MOS.

\add{
We recruited 50 gender-balanced, English-speaking participants over the age of 18 via the Internet using the Prolific crowdsourcing system~\cite{prolific}. We used a javascript-based MUSHRA testing tool~\cite{mushraJS} for online evaluation (Figure~\ref{fig:mushraweb}, the system is available from \url{https://github.com/rkmt/mushraJS_prolific} ). We used the same whisper-normal voice samples as in the previous MOS test. The participants' responses are collected over the Internet. 
}

\begin{figure}
    \centering
    \includegraphics[width=0.8\linewidth]{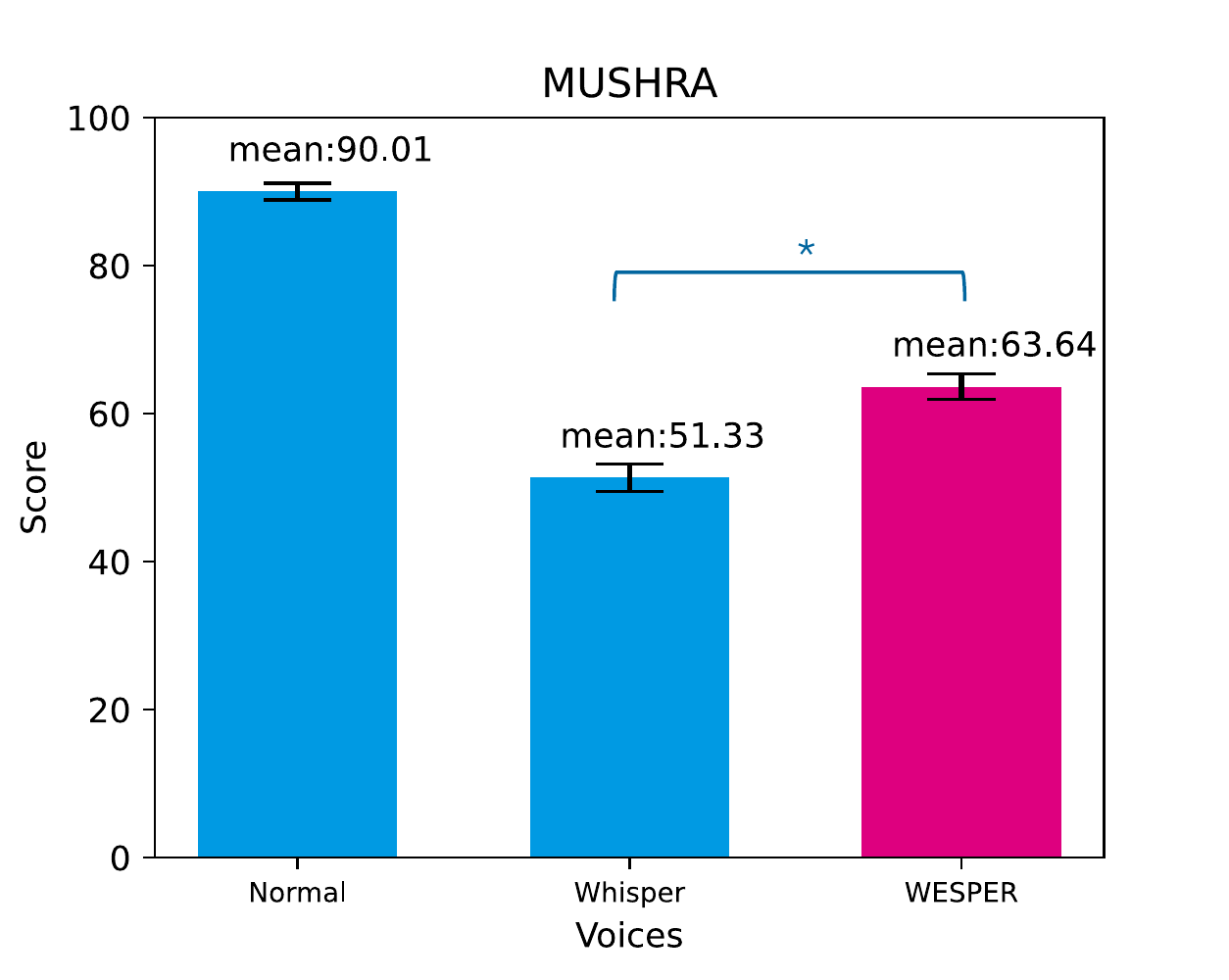}
    \caption{Speech quality evaluation between whispered and WESPER converted voices by MUSHRA (MUltiple Stimuli with Hidden Reference and Anchor). ({\tt *}: $p < 0.01$, t-test)}
    \Description{Speech quality evaluation between whispered and WESPER converted voices by MUSHRA (MUltiple Stimuli with Hidden Reference and Anchor). ({\tt *}: $p < 0.01$, t-test)}
    \label{fig:mushra}
\end{figure}

The results are presented in Figure~\ref{fig:mushra}. It revalidates the results of the MOS evaluation ($p < 0.01$ through pairwise t-test, effect size = 0.57).

According to these evaluations, we can draw the following conclusions:
\begin{itemize}
\item WESPER can convert whispered voices to normal voices.
\item Speechdata converted with WESPER had a better MOS than the whispered source voice.
\item WESPER preserved the natural prosody of the source whispered voice.
\end{itemize}

\subsection{Speech Recognition Accuracy}

There are two possible ways to use WESPER as a speech recognizer, including speech recognition with WESPER and recognition of speech converted with WESPER by other speech recognizers.

In the former method, a WESPER model (based on HuBERT) pre-trained with whispered and normal speech is fine-tuned using the whispered corpus, and text is inferred from the whispered voice.
In the latter case, the whispered speech converted by WESPER can control any other speech-enabled device. 
If the whispered speech can be converted to normal speech, then the existing speech-enabled devices can be used immediately without the need to modify them for whispered speech.

Using the existing speech recognizer (google
cloud speech-to-text~\cite{googlespeech}) as a reference, we measured the speech recognition accuracy of normal speech, whispered speech, and whispered speech converted by WESPER.

The wTIMIT corpus was used for the measurements. wTIMIT is a TIMIT-compliant transcription containing both normal and whispered speech with labels. This corpus evaluated the recognition accuracy of whispered speech converted by WESPER using the recognition accuracy of normal and whispered speech as the baseline.

The results are summarized in Table~\ref{tab:wer} in terms of word error rate (WER) and character error rate (CER), as well as bilingual evaluation understudy (BLEU).
As shown in the table, the recognition accuracy was not high when whispered speech was directly recognized (WER=44.70\%), but the accuracy improved (WER=26.68\%) after conversion with WESPER. 

When tested with wTIMIT(W), HuBERT-base pre-trained with Librispeech and wTIMIT(N,W) shows better results than the google cloud speech-to-text (HuBERT: WER=33.06\%, Google: WER=44.70\%). It is speculated that the effect of pre-training with a mixture of whispered and normal speech (but without fine-tuning with whispered voice) may contribute to the accuracy of ASR.

Notably, WESPER is not trained with a labeled corpus. It is pre-trained with whispered and normal speeches, which improves the speech recognition accuracy.

Therefore, speech recognition via WESPER conversion does not require the preparation of a corpus of whispered speech for specific or unspecified speakers.

\begin{figure}
    \centering
    \includegraphics[width=0.8\linewidth]{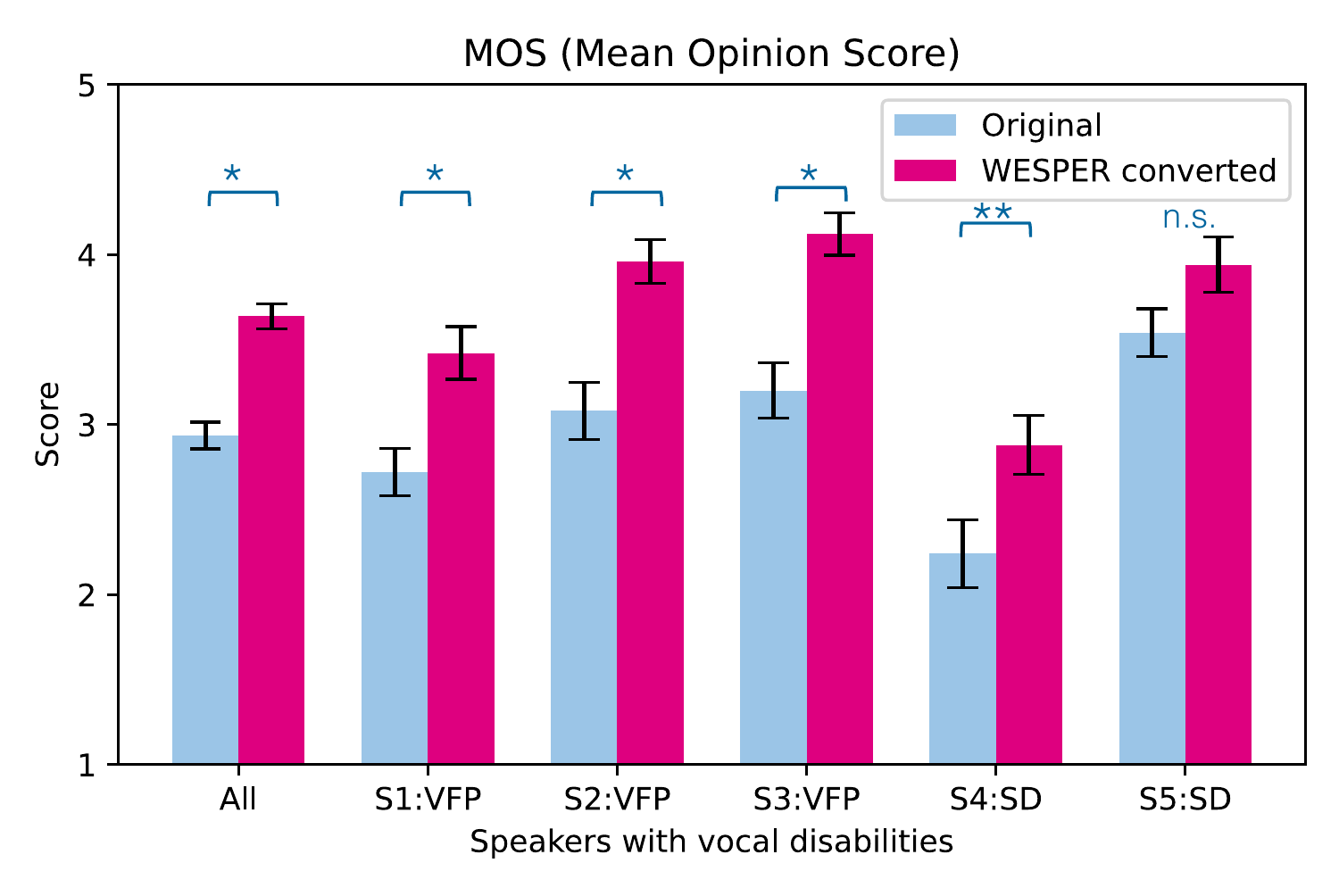}
    \caption{Speech quality evaluation for people with speech disorders ranked by 5-point MOS: ({\tt *}: $p < 0.01$, {\tt **}: $p < 0.05$, {\tt S1-S5}: speakers, {\tt VFP}: Vocal Fold Polyps, {\tt SD}: Spasmodic Dysphonia)}
    \Description{Speech quality evaluation for people with speech disorders ranked by 5-point MOS: ({\tt *}: $p < 0.01$, {\tt **}: $p < 0.05$, {\tt S1-S5}: speakers, {\tt VFP}: Vocal Fold Polyps, {\tt SD}: Spasmodic Dysphonia)}
    \label{fig:dismos}
\end{figure}

\begin{figure}
    \centering
        \includegraphics[width=0.8\linewidth]{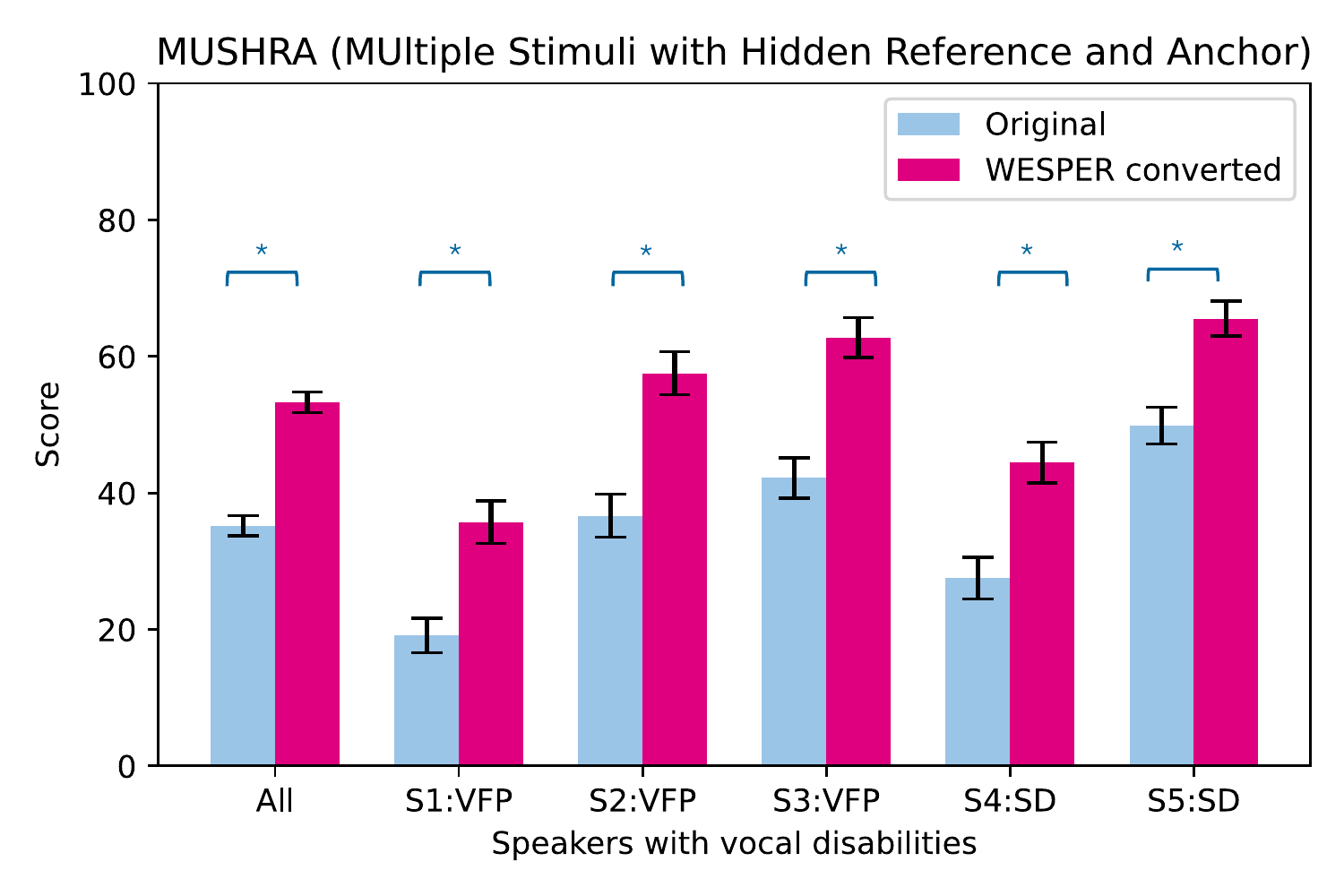}
    \caption{Speech quality evaluation for people with vocal disabilities ranked by MUSHRA.  ({\tt *}: $p < 0.01$, {\tt S1-S5}: speakers, {\tt VFP}: Vocal Fold Polyps, {\tt SD}: Spasmodic Dysphonia)}
    \Description{Speech quality evaluation for people with vocal disabilities ranked by MUSHRA.  ({\tt *}: $p < 0.01$, {\tt S1-S5}: speakers, {\tt VFP}: Vocal Fold Polyps, {\tt SD}: Spasmodic Dysphonia)}
    \label{fig:dis_mushra}
\end{figure}

\begin{figure}
    \centering
    \includegraphics[width=0.8\linewidth]{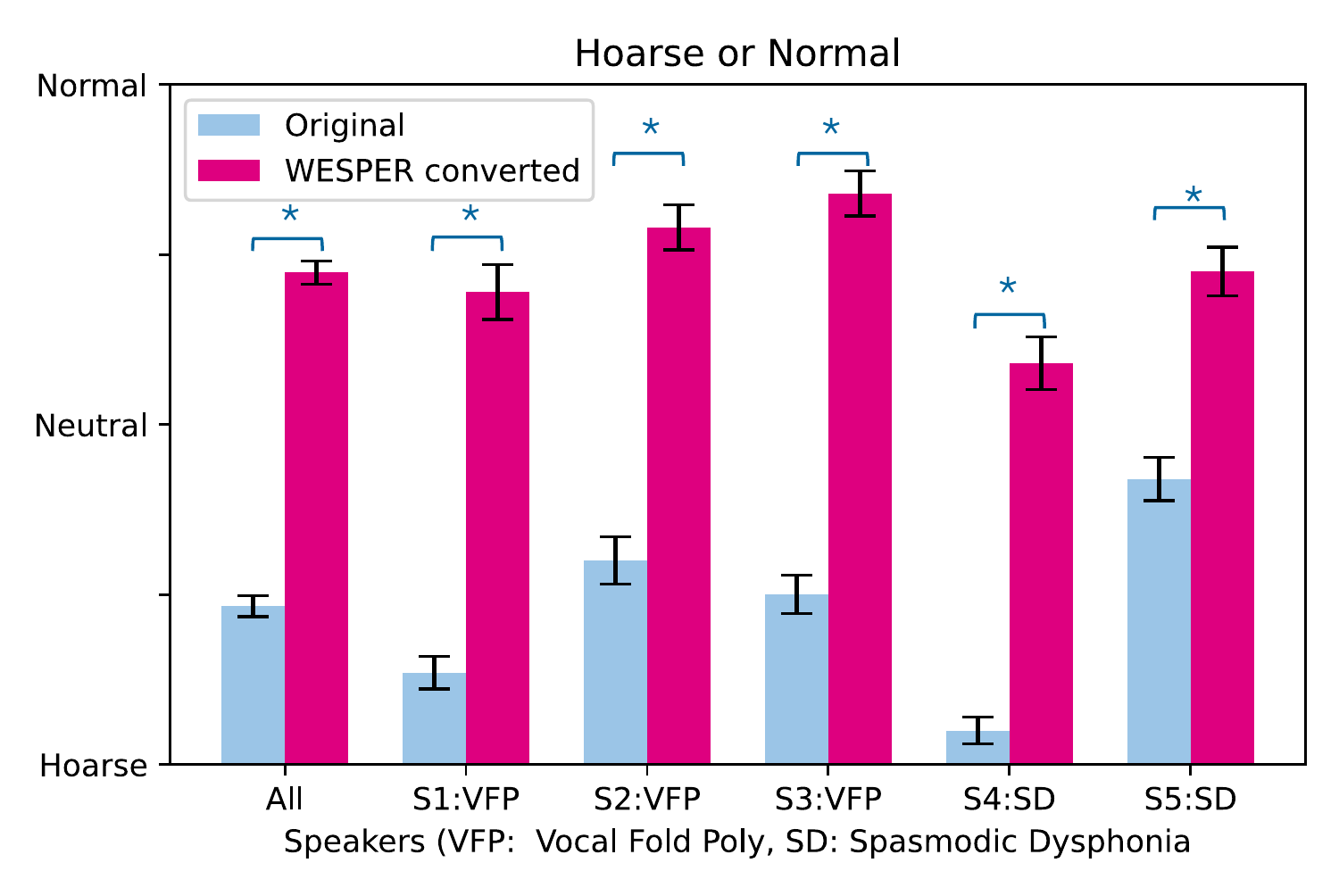}
    \caption{Hoarse-Normal assessments of utterances by people with speech disorders ({\tt *}: $p < 0.01$).}
    \Description{Hoarse-Normal assessments of utterances by people with speech disorders ({\tt *}: $p < 0.01$).}
    \label{fig:disHoarse}
\end{figure}

\begin{figure}
    \centering
    \includegraphics[width=0.8\linewidth]{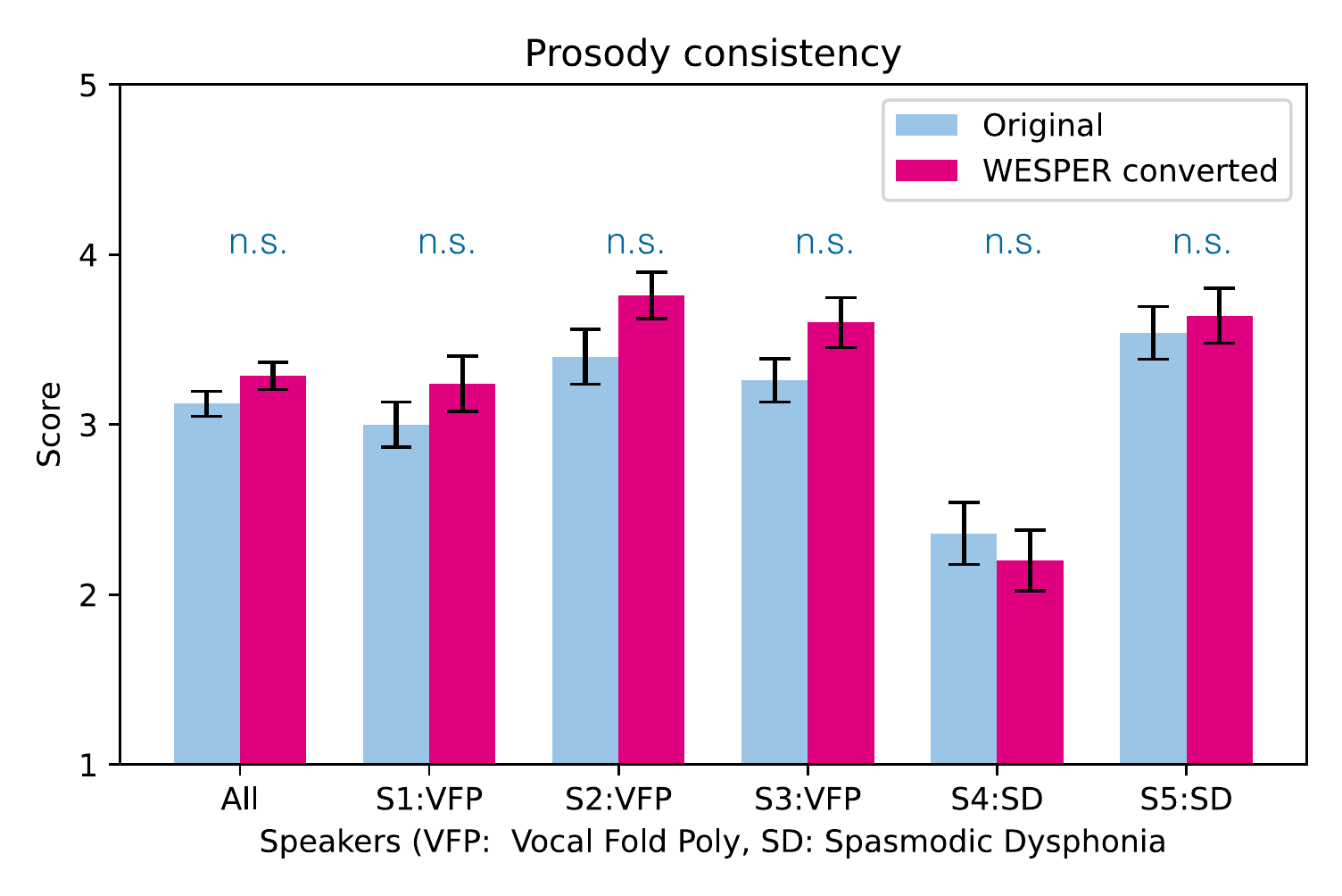}
    \caption{Speech prosody consistency of utterances by people with speech disorders. ({\tt n.s.}: not significant)}
    \Description{Speech prosody consistency of utterances by people with speech disorders. ({\tt n.s.}: not significant)}
    \label{fig:disPros}
\end{figure}

\subsection{Evaluations on Speech Reconstruction for People with Speech Disorders}

An important goal of WESPER is the reconstruction of atypical speech of people with speech disorders or hearing impairments. This makes their speech more understandable to people who are not familiar with their individual speech patterns. 

Dysphonia is an involuntary hoarse, breathy, or strained sounding voice or low volume or pitch. There are various causes, including spasms, polyps of the vocal tract, and other causes. If the vocal cords have been removed because of throat cancer or other causes, the voice becomes extremely difficult to produce. For hearing impairment, even if the vocal cords are not affected, the control of the vocal cords becomes difficult, resulting in dysphonia. 
Electrolarynx are used to mechanically vibrate the throat as a vocalization for people with vocal cord damage, but the sound produced is artificial, the pitch conversion is deterministic, and there is a large deviation from normal vocalization.
The communication deficit caused by dysphonia is a serious problem, and its elimination by voice conversion technology should be of great social value.

To investigate the quality of the improvement by WESPER voice conversion, we evaluated the speech utterances of people with two types of speech disorders, as described below. 

\begin{description}
\item[Vocal Fold Polyps (we will refer as VFPs): ] VFPs are among the most common benign lesions of the larynx, affecting the quality of voice production.
\item[Spasmodic Dysphonia (we will refer as SD): ] This is also called laryngeal dystonia, SD is another common neurological disorder that affects voice and speech. It is a lifelong condition that causes spasms of the muscles that  produce the voice. 
\end{description}

We used the Saarbruecken Voice Database (SVD) corpus~\cite{SVD,svd2} to sample utterances from people with VFP and SD. 
The SVD is a commonly used corpus of voices of people with speech disorders. It contains recordings of vowel utterances and a recording of the German reference sentence ``Guten Morgen, wie geht es Ihnen?'' (``Good morning, how are you?''). 
This sentence was used for evaluation.

Fifty participants were recruited for the evaluation using the Prolific~\cite{prolific}. The recruited participants were evenly balanced in terms of gender and were all over the age of 18. The participants were fluent in German in addition to English, as the reference sentence used was in German.

The results are presented in Figures~\ref{fig:dismos}, \ref{fig:dis_mushra}, \ref{fig:disHoarse}, and \ref{fig:disPros}.
Figure~\ref{fig:dismos} shows the results in terms of MOS. For both SD and VFP, WESPER-converted voices had higher MOS scores ($p < 0.01$, effect size=0.66) and MUSHRA scores ($p < 0.01$, effect size=0.79).
Figure~\ref{fig:disHoarse} shows the responses to the question ``Is this voice hoarse or normal?'' and there is a clear improvement in the WESPER-converted voices compared to the original VFP and SD voices ($p < 0.01$).
Figure~\ref{fig:disPros} shows the answers to the question ``Does the voice use consistent articulation, standard intonation, and prosody?''. Here, WESPER-converted and original voices showed nearly equal scores, although WESPER-converted voices showed slightly better scores. It can be assumed that WESPER did not affect the prosody of the original speech.

According to these evaluations, we can make the following conclusions.
\begin{itemize}
\item WESPER-converted voices of people with VFP and SD speech disabilities showed better quality. This suggests that WESPER can improve the quality of the speech of people with these conditions in terms of its intelligibility by people unfamiliar with their individual speech patterns.
\item Similarly, WESPER can improve the naturalness of original VFP and SD speech.
\item WESPER could also preserve the natural prosody of the source speech.
\end{itemize}

Notably, this test was performed on German sentences. Although the WESPER pre-training was performed only on English speech and not on German speech, the prosody of the source speech was preserved and its MOSs were improved. Therefore, this result may demonstrate the language-independence ability of the WESPER model. The attached video shows an example of whisper to normal conversion in Japanese.
Because wav2vec 2.0, the base model for STU, has also shown language-independent 
pre-training performance~\cite{https://doi.org/10.48550/arxiv.2006.13979}, the language-independent ability of WESPER could be a feature worth evaluating in the future.

\subsection{Speech Reconstruction Evaluation for People with Hearing Impairment}

\begin{figure}
    \centering
    \includegraphics[width=0.8\linewidth]{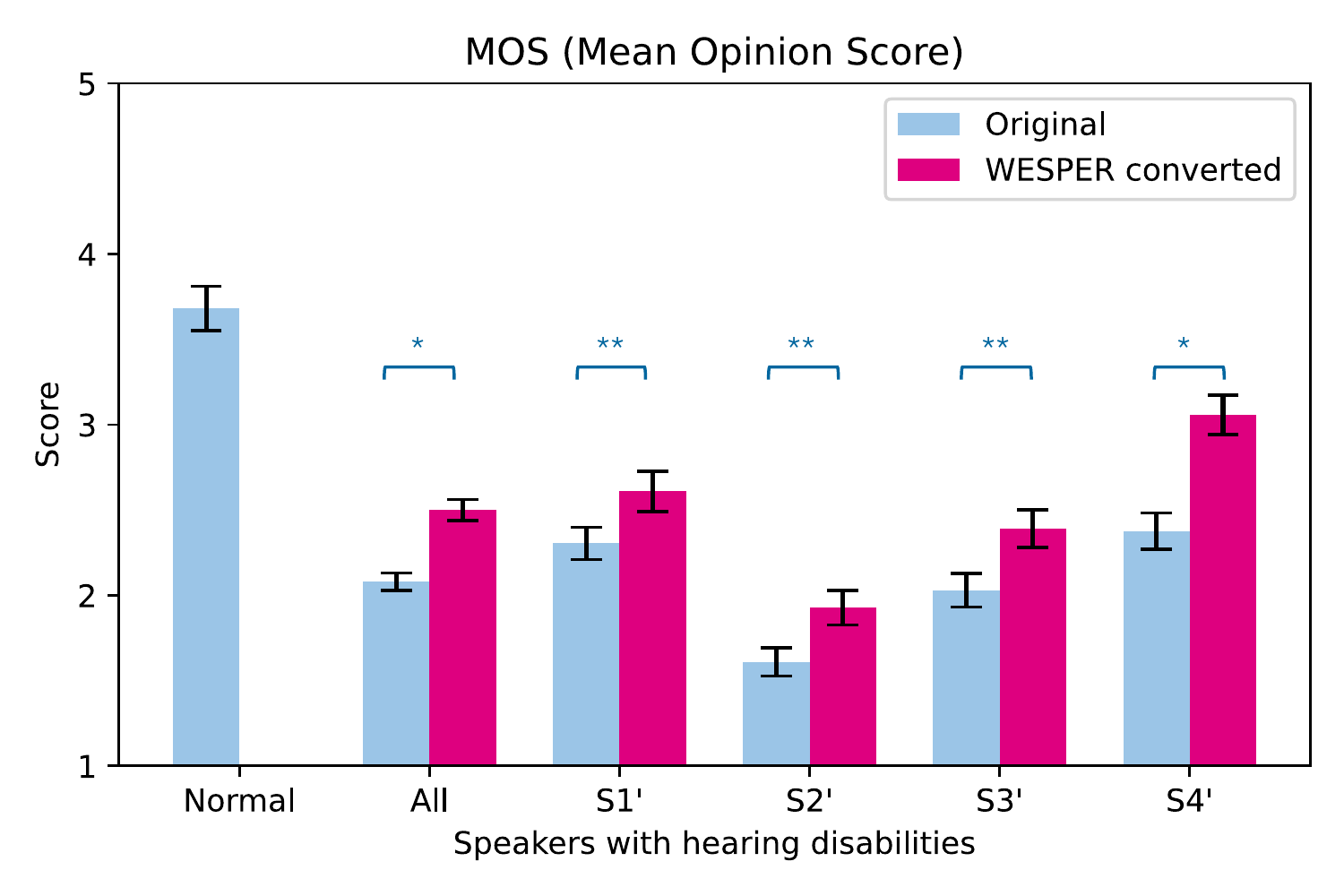}
    \caption{Speech quality evaluation for people with hearing impairment: Speech quality was ranked by 5-point MOS. ({\tt S1'-S4'} :speakers,  {\tt *}: $p < 0.01$, {\tt **}: $p < 0.05$)}
   \Description{Speech quality evaluation for people with hearing impairment: Speech quality was ranked by 5-point MOS. ({\tt S1'-S4'} :speakers,  {\tt *}: $p < 0.01$, {\tt **}: $p < 0.05$)}
    \label{fig:HDmos}
\end{figure}

\begin{figure}
    \centering
    \includegraphics[width=0.8\linewidth]{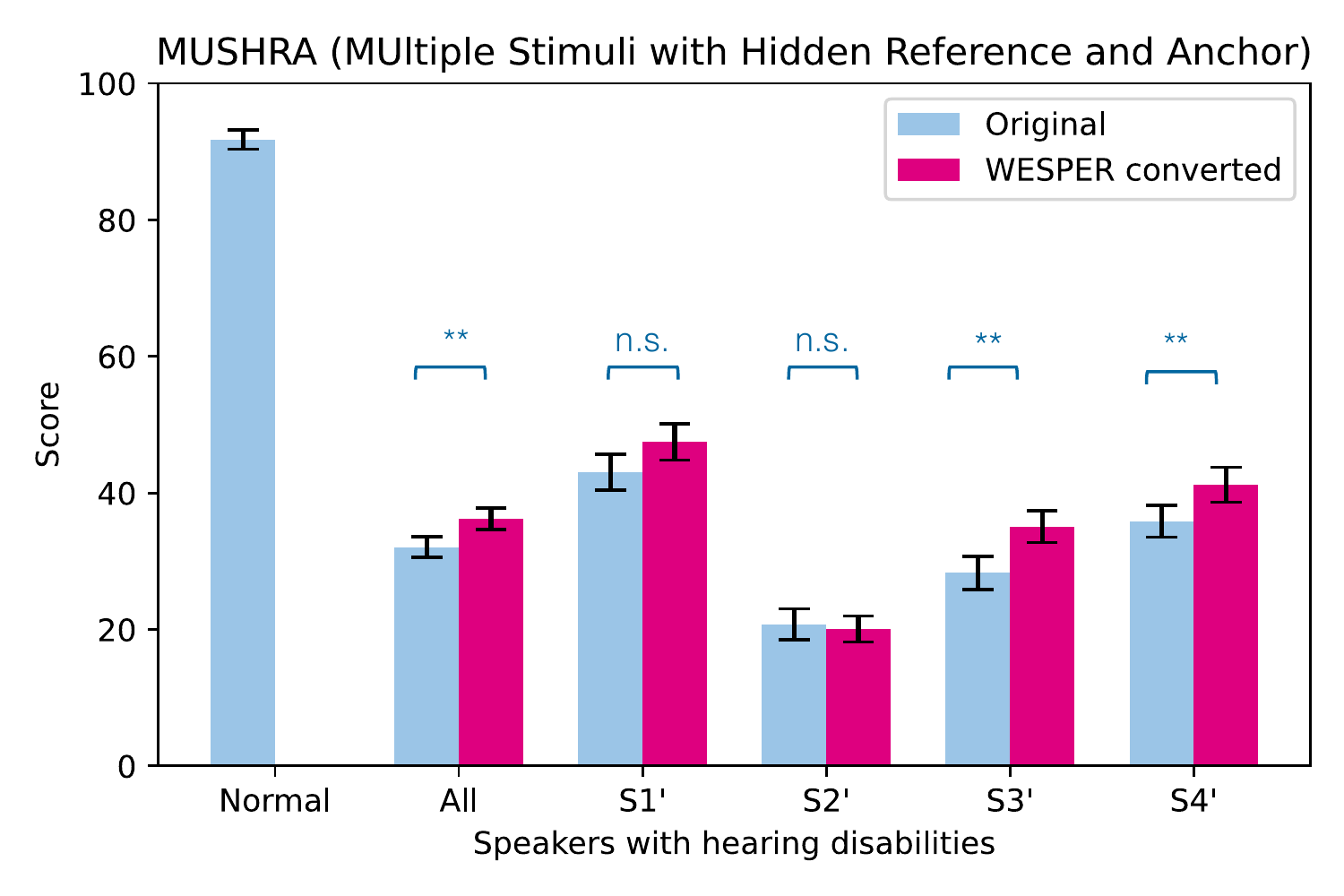}
    \caption{Speech quality evaluation for people with hearing impairment: Speech quality was ranked by MUSHRA. ({\tt S1'-S4'}: speakers, {\tt **}: $p < 0.05$)}
    \Description{Speech quality evaluation for people with hearing impairment: Speech quality was ranked by MUSHRA. ({\tt S1'-S4'}: speakers, {\tt **}: $p < 0.05$)}
    \label{fig:HD_mushra}
\end{figure}

\begin{figure}
    \centering
    \includegraphics[width=0.8\linewidth]{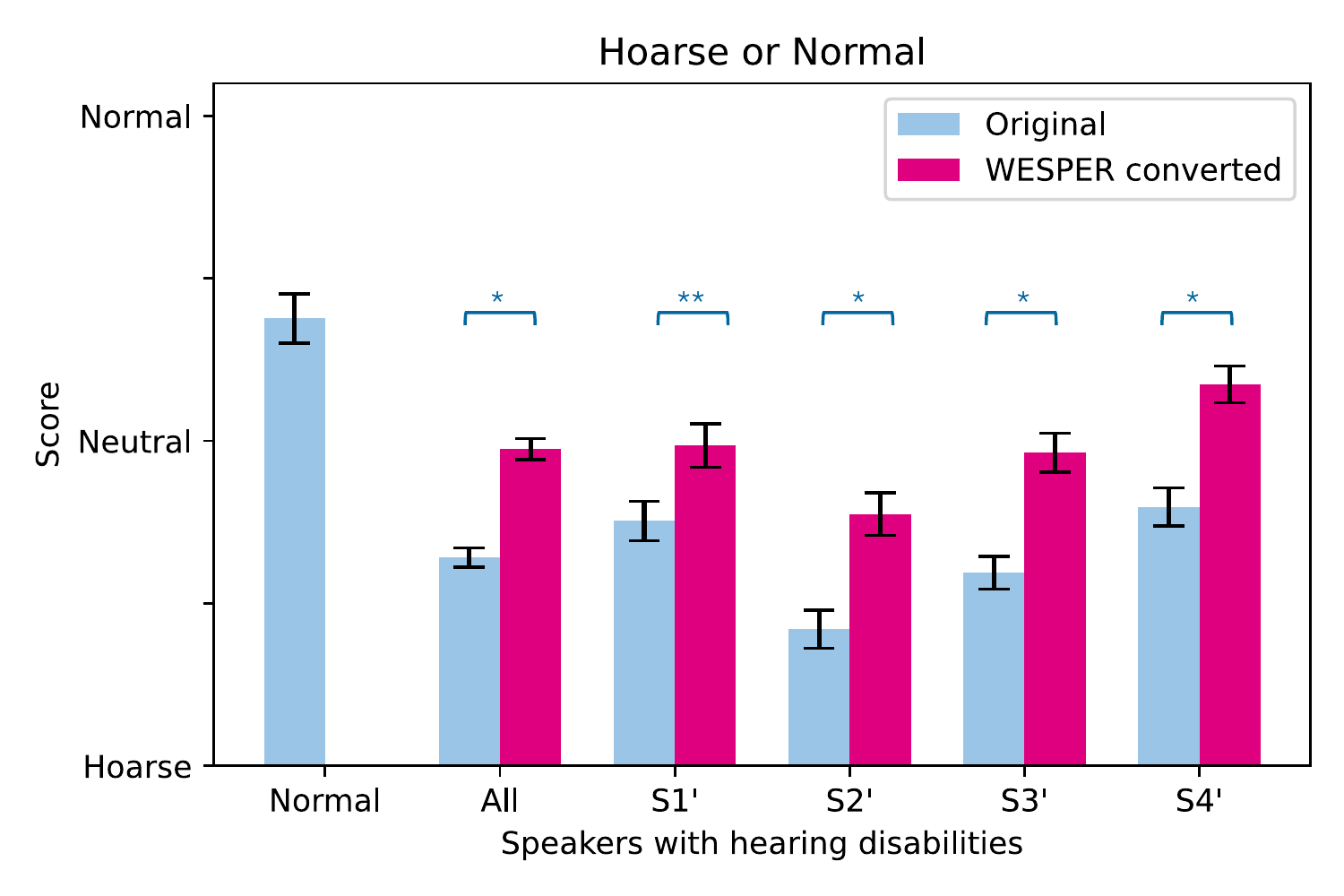}
    \caption{Hoarse-Normal assessments of utterances by people with hearing impairment. ({\tt *}: $p < 0.01$, {\tt **}: $p < 0.05$)}
    \Description{Hoarse-Normal assessments of utterances by people with hearing impairment. ({\tt *}: $p < 0.01$, {\tt **}: $p < 0.05$)}
    \label{fig:HDhoarse}
\end{figure}

\begin{figure}
    \centering
    \includegraphics[width=0.8\linewidth]{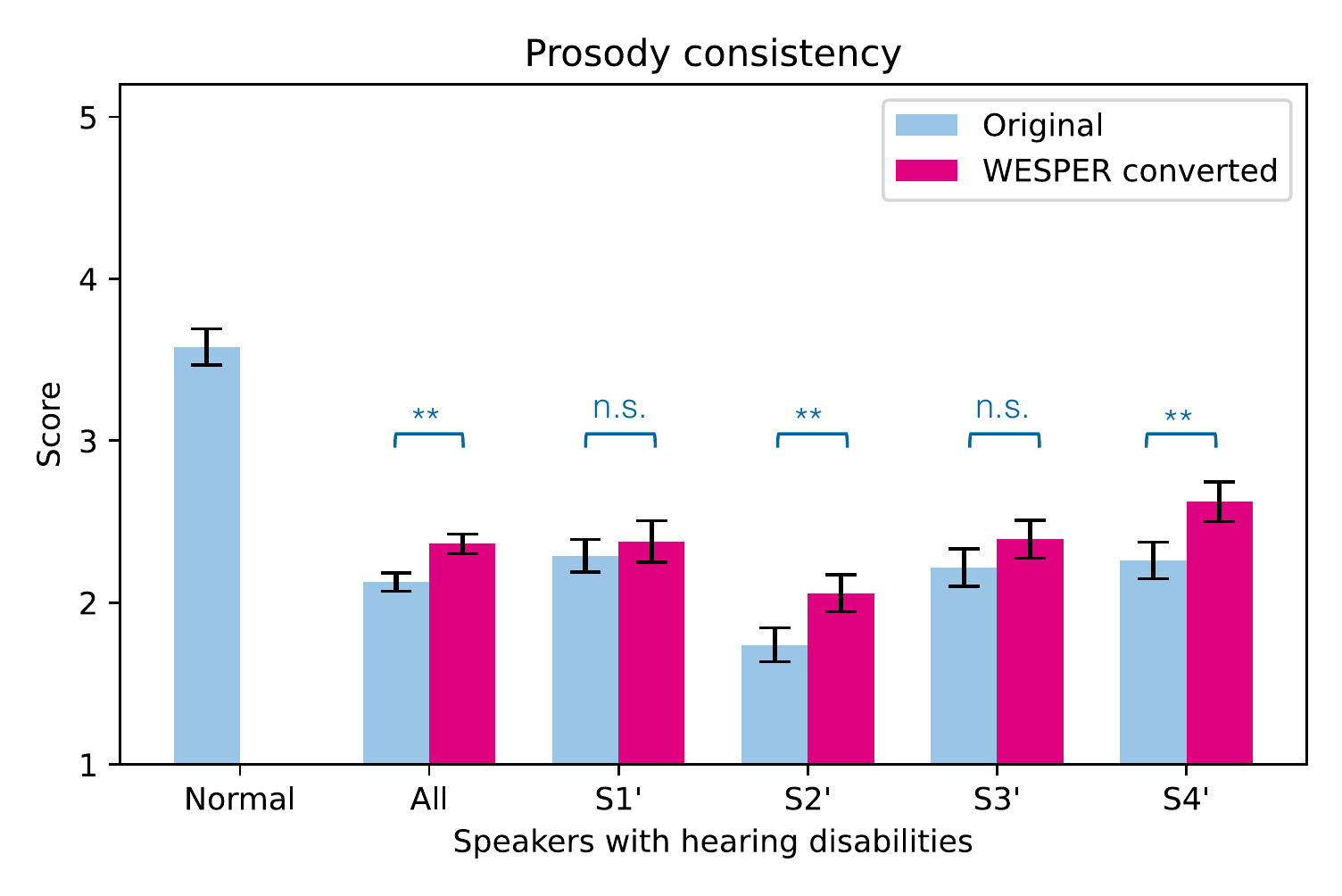}
    \caption{Speech natural prosody consistency by people with hearing impairment. ({\tt **}: $p < 0.05$)}
    \Description{Speech natural prosody consistency by people with hearing impairment. ({\tt **}: $p < 0.05$)}
    \label{fig:HDprosody}
\end{figure}

Finally, we evaluated the effect of speech reconstruction by WESPER for people with hearing impairments. 
People with hearing loss cannot hear their own speech and that of others; therefore, they tend to have difficulty speaking in a way that is easily understood by general speakers.
However, because their vocal organs are normal, their speech has different characteristics from that of people with dysphasia.

We used the ``corpus of deaf speech for acoustic and speech production research''~\cite{deafspeech}. Utterances of five speakers (one was a hearing speaker and the other four were hearing impaired) were used. We recruited 50 evaluation participants using Prolific~\cite{prolific}, who were evenly balanced in terms of gender, fluent English speakers, and over the age of 18. We asked the participants to perform a 5-point ranking of each utterance in terms of MOS, hoarse-normal, and natural prosody.    

The results are presented in Figures~\ref{fig:HDmos}, \ref{fig:HD_mushra}, \ref{fig:HDhoarse}, and \ref{fig:HDprosody}.
These results indicate that MOS, MUSHRA, and other rating scores were higher for those converted by WESPER ($p < 0.05$, effect size = 0.45 for MOS, $p < 0.05$, effect size = 0.21 for MUSHRA), but the degree of improvement was less than for the speakers with voice disorders. In addition, prosody ratings were significantly lower for the speech of the hearing impaired compared to the speech of the hearing speakers. Therefore, these results indicate that people with hearing loss may have difficulty controlling prosody while speaking.

The results of all experiments are summarized in Table~\ref{tab:questions}.

\begin{table}
    \centering
    \begin{tabular}{c|c|cccc}
    \toprule
    \multicolumn{2}{c|}{Speaker Type} & \multicolumn{1}{c}{MOS} & \multicolumn{1}{c}{MUSHRA} & \multicolumn{1}{c}{Q1} & \multicolumn{1}{c}{Q2}\\
    \midrule
    \multicolumn{2}{c|}{Normal}              & $4.15$ & $90.01$ & $4.16$ & $3.99$\\
    \multicolumn{2}{c|}{Whisper}             & $3.27$ & $51.33$ & $2.90$ & $3.31$\\
    \multicolumn{2}{c|}{\bf WESPER[Whisper]} & $3.79$ & $63.64$ & $3.97$ & $3.43$\\
    \midrule
    \multicolumn{5}{c}{Speakers with speech disorders} \\ \hline
    \multirow{6}{*}{VFP} & S1            & $2.72$ & $19.13$ & $1.54$ & $3.00$ \\
        & WESPER[S1]                     & $3.42$ & $35.72$ & $3.78$ & $3.24$ \\
    & S2                                 & $3.08$ & $36.66$ & $2.20$ & $3.40$ \\
        & WESPER[S2]                     & $3.96$ & $57.53$ & $4.16$ & $3.76$ \\
        & S3                             & $3.20$ & $42.19$ & $2.00$ & $3.26$ \\
                & WESPER[S3]             & $4.12$ & $62.74$ & $4.36$ & $3.60$ \\
        \hline
    \multirow{4}{*}{SD} & S4             & $2.24$ & $27.51$ & $1.20$ & $2.36$ \\    
        & WESPER[S4]                     & $2.88$ & $44.45$ & $3.36$ & $2.20$ \\
        & S5                             & $3.54$ & $49.86$ & $2.68$ & $3.54$ \\       
                 & WESPER[S4]            & $3.94$ & $65.53$ & $3.90$ & $3.64$ \\      
        \hline
        \multicolumn{2}{c|}{All}         & $2.93$ & $35.20$ & $1.93$ & $3.12$ \\
        \multicolumn{2}{c|}{WESPER[All]} & $3.64$ & $53.30$ & $3.89$ & $3.29$ \\
        \midrule
        \multicolumn{5}{c}{Speakers with hearing impaired} \\
        \hline
       \multicolumn{2}{c|}{Normal}       & $3.68$ & $91.75$ & $3.76$ & $3.75$\\
       \multicolumn{2}{c|}{S1'}          & $2.30$ & $43.05$ & $2.29$ & $2.51$\\
        \multicolumn{2}{c|}{WESPER[S1']} & $2.61$ & $47.47$ & $2.38$ & $2.97$\\
       \multicolumn{2}{c|}{S2'}          & $1.61$ & $20.76$ & $1.74$ & $1.84$\\
        \multicolumn{2}{c|}{WESPER[S2']} & $1.93$ & $20.05$ & $2.06$ & $2.55$\\
       \multicolumn{2}{c|}{S3'}          & $2.03$ & $28.30$ & $2.19$ & $2.22$\\
        \multicolumn{2}{c|}{WESPER[S3']} & $2.39$ & $35.07$ & $2.39$ & $2.93$\\        
       \multicolumn{2}{c|}{S4'}          & $2.38$ & $35.86$ & $2.26$ & $2.59$\\
        \multicolumn{2}{c|}{WESPER[S4']} & $3.06$ & $41.23$ & $2.62$ & $3.35$\\ 
                \hline
        \multicolumn{2}{c|}{All}         &$2.08$ & $32.07$ & $2.28$ & $2.12$ \\
       \multicolumn{2}{c|}{WESPER[All]}  &$2.50$ & $36.21$ & $2.94$ & $2.36$ \\
                
        \bottomrule
    \end{tabular}

    {\small
    \begin{tabular}{ll}
    MOS & Mean Opinion Score \\
    MUSHRA & MUltiple Stimuli with Hidden Reference and Anchor\\
Q1 &  {\it `Is this voice hoarse or normal?'} \\
    Q2 & \mmm{\it `Is the voice use consistent articulation, \\standard intonation and prosody?'} \\
    WESPER[$\bullet$] & WESPER converted $\bullet$ \\
    VFP &  Vocal Fold Polyps speakers \\
SD & Spasmodic Dysphonia speakers \\
S1-S5, S1'-S4' & speaker IDs \\
    \end{tabular}
    }
    
\caption{Summary of Voice Conversion Quality Evaluation (MUSHRA: 100-point score, others:5-point score)}
\Description{Summary of Voice Conversion Quality Evaluation (MUSHRA: 100-point score, others:5-point score)}
    \label{tab:questions}
\end{table}

\section{Discussions}

\subsubsection*{Combination of WESPER and User-Dependent Fine-tuning}

The results of the evaluation experiments indicated that the degree of improvement in the speech of people with hearing impairments was less than that of the speech of people with dysphasia. Therefore, we assume that the former has significant prosodic variations. Although the primary purpose of this study was to achieve speaker-independent speech conversion by pre-training alone, we plan to conduct additional experiments to investigate whether the conversion performance can be improved by applying small-scale fine-tuning to each speaker.

Even in this case, we believe that speech pairs are necessary, but text transcription is not required. A pair of whispered and hoarse speech utterances and normal speech converted to a common speech unit by STU can be used instead of text transcription, as in the case of ASR fine-tuning.

\subsubsection*{Audio Input Device Suitable for Whispered Speech}

As demonstrated in this study, whispered or hoarse voices can be converted into normal voices. 
In practice, the selection of an appropriate audio input device would be important.
We are currently testing our proposed approach with normal headsets and a directional array microphone (designed for smart speakers), and have obtained good results. For wearable devices, a non-audible murmur (NAM) microphone that detects skin vibrations could be used for inaudible utterances~\cite{NAM2005}. Combining with noise reduction techniques such as ~\cite{Chatterjee_2022,https://doi.org/10.48550/arxiv.2210.15324} is another important future direction.

Philips and Dyson are also developing masks that provide powered respiratory ventilation to protect against air pollution and infectious diseases~\cite{pmask,dysonzone}. A microphone can be placed inside such masks 
to pick up a whispered voice, which would give an effect almost equivalent to silent speech.

\subsubsection*{Human-AI Integration}

This research concerns machine-learning techniques for converting the whispered speech of unspecified speakers into normal speech. In practice, however, we established that users noticed that some similar whispered utterances were easily converted to normal speech, whereas others were difficult. They tried to speak by relying on machine learning on the user side. This interaction suggests that machine learning does not only unilaterally extend human capabilities, but that further synergistic effects can be achieved by learning on the human side as well. This can be seen as an actual example of human-AI integration~\cite{uist19rekimoto} through vocalization.

\section{Conclusion}
In this study, we proposed WESPER as a mechanism for the real-time conversion of whispered speech into normal speech. We confirmed that a common speech unit can be obtained by self-supervised learning even when the acoustic features of whispered and normal speech are different. Speech reproduction can be learned from speech data of arbitrary target speakers only without using text labels.
There is no need to train for each user, nor is there a need for parallel data for whispered and normal speech. From speech units, WESPER can reconstruct utterances of any target speaker and requires only unlabeled speech data from target speakers. We confirmed that the quality of the converted speech is improved and that the prosody of the speech is preserved. Additionally, we reported an evaluation of the results of reconstructing speech utterances of people with speech disorders and hearing impairments.

\begin{acks}
We are grateful to the reviewers for their valuable comments.
This work was supported by JST Moonshot R\&D Grant Number JPMJMS2012, JST CREST Grant Number JPMJCR17A3, the commissioned research by NICT Japan, and The University of Tokyo Human Augmentation Research Initiative.
\end{acks}

\bibliographystyle{ACM-Reference-Format}
\bibliography{main}

\end{document}